\documentclass[english,aps,prl,twocolumn,amsmath,amssymb,showpacs,notitlepage,longbibliography]{revtex4-1}
\usepackage{bm}
\usepackage{hhline}
\usepackage{bbold}
\usepackage{makecell}
\usepackage{amsmath}
\usepackage{amssymb}
\usepackage{mathdots}
\usepackage{tabularx}
\usepackage{graphicx}
\usepackage{hyperref}
\usepackage{bbding}
\usepackage{tikz}
\usepackage{bbding} 
\usepackage{braket}
\usepackage{bm}
\usepackage{leftindex} 
\usepackage{siunitx}
\usepackage{multirow}

\makeatletter\def\fnum@figure{\textbf{Fig.~\thefigure}} \makeatother
\makeatletter\def\fnum@table{\textbf{Table~\thetable}} \makeatother

\usepackage{babel}

\begin{document}
\title{Heesch Nodal Lines in Inadmissible Achiral Antiferromagnets} 
\author{Xing-Yao Guo} 
\author{Chung-Yuen Chan}
\author{Zi-Ting Sun}
\thanks{zsunaw@connect.ust.hk}
\author{Kam Tuen Law} 
\thanks{phlaw@ust.hk}
\affiliation{Department of Physics, Hong Kong University of Science and Technology, Clear Water Bay, Hong Kong, China} 

\date{\today}
\begin{abstract} 
    Recently, a new class of Weyl semimetals in antiferromagnets named Heesch Weyl semimetals was discovered, which have inadmissible chiral magnetic point group symmetries (inadmissible magnetic point groups are incompatible with ferromagnetic order) and distinctive surface Fermi arcs.
    In Heesch Weyl semimetals, the Weyl points are pinned at high symmetry momenta with two-dimensional irreducible corepresentations in the Brillouin zone. 
    As the Weyl points are pinned, the Weyl points with opposite topological charges cannot emerge or be brought together for creation and annihilation as in conventional Weyl semimetals. 
    In this work, we show that when mirror or rotoinversion symmetries are restored so that the point group becomes achiral, long doubly degenerate lines connecting Weyl points with opposite topological charges emerge.
    We call these lines the Heesch nodal lines (HNLs) and their host materials the Heesch nodal line antiferromagnets.
    HNLs result in a large number of two-dimensional massless Dirac cones for planes intercepting the HNLs in the Brillouin zone.  
    Moreover, a large subset of the HNL antiferromagnets has the special property that the lowest nonvanishing order of the nonlinear anomalous Hall effect starts with the third order.  
    First-principles calculations on representative collinear and noncollinear antiferromagnets, such as MnTe, CrSb, and Mn$_3$GaN, confirm our predictions on the presence of HNLs. 
    When the inadmissible symmetry is broken by strain, the double degeneracy of the HNLs is lifted and the associated massless Dirac cones are gapped out, providing a route to realizing sizable anomalous Hall effects in antiferromagnetic crystals.
    We conclude that all inadmissible antiferromagnets without parity-time symmetry are topological. They are either Heesch Weyl antiferromagnets or Heesch nodal line antiferromagnets.
\end{abstract}
\maketitle

\section{Introduction} 
Topological phases of matter have become a central paradigm in modern condensed matter physics, extending from gapped systems such as topological insulators to gapless topological semimetals \cite{Hasan_review_2010,  Qi_review_2011, Armitage_review_2018}, including Dirac semimetals \cite{PhysRevLett.108.140405, PhysRevB.85.195320, PhysRevB.88.125427, PhysRevLett.113.027603, 10.1126/science.1245085, Liu2014, Yang2014, 10.1126/science.aac6089}, Weyl semimetals \cite{PhysRevB.83.205101, PhysRevLett.107.127205, PhysRevLett.107.186806, PhysRevB.84.075129, PhysRevX.5.011029, PhysRevX.5.031013, Xu2015, Wang2018, Liu2019, Noam2019}, and nodal line metals \cite{PhysRevB.84.235126, PhysRevB.90.205136, PhysRevB.92.045108, PhysRevLett.115.036806, PhysRevB.92.081201, Bian2016, Schoop2016, Wang2017, fu2019, Ilya2019, Yang2021, Knoll2022, Zhuang2026}.
In particular, Weyl semimetals are characterized by isolated twofold band crossings with a linear dispersion near each Weyl node. 
These Weyl nodes carry quantized topological charges and appear in pairs of opposite charges.  
The chiral topology of Weyl nodes gives rise to surface Fermi arcs and a variety of exotic transport and optical phenomena, such as the chiral anomaly \cite{Nielsen1983, Son2013, Huang2015, Zhang2016}, nonlinear optical responses \cite{Chan2016, deJuan2017, Wu2017}, and magneto-optical effects \cite{Ashby2013, Kargarian2015, Okamura2020}.   

 \begin{figure}[!t]
	\centering
	\includegraphics[width=0.98\columnwidth]{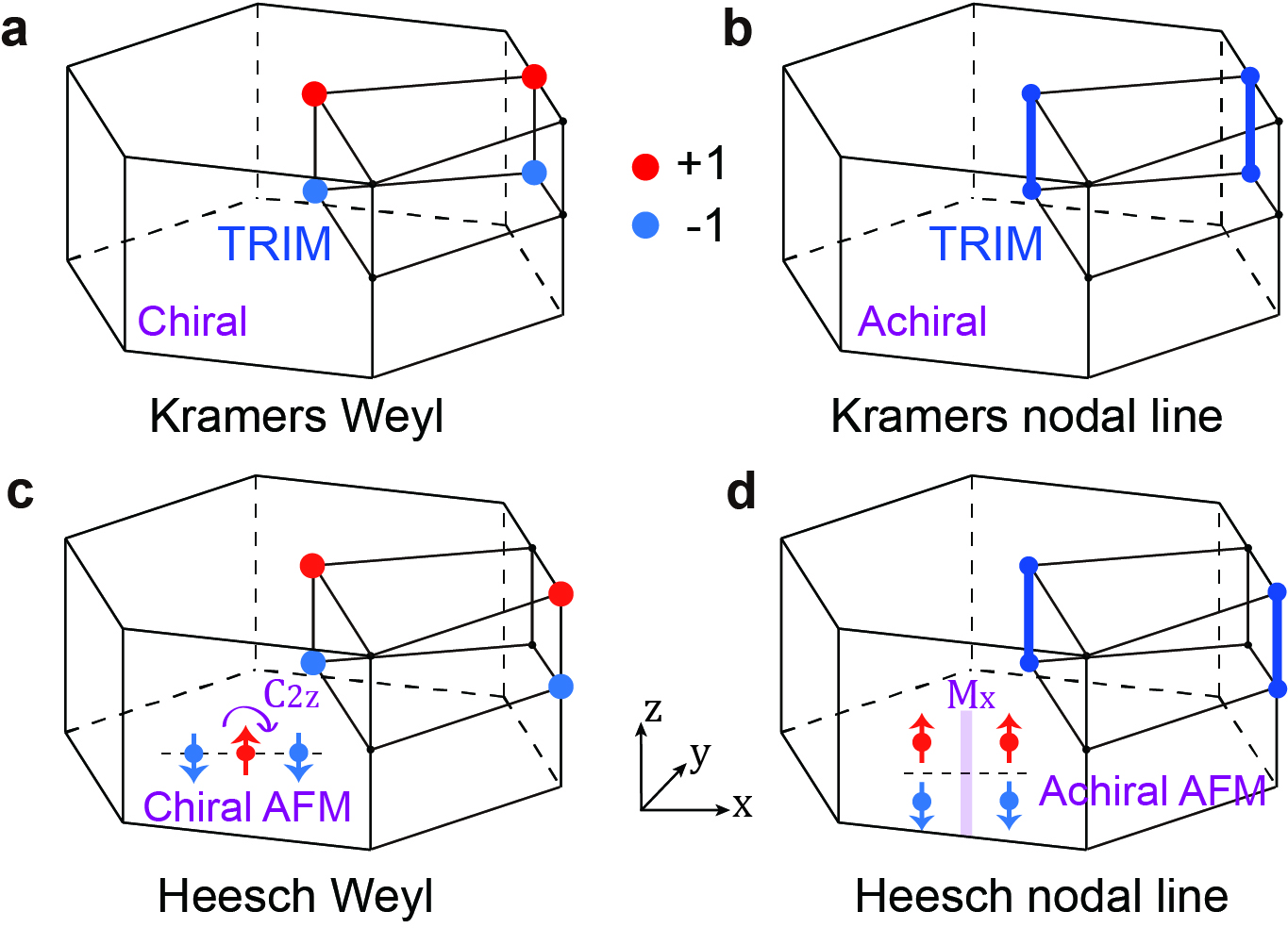}
	\caption{\textbf{Four types of topological metals.}
		\textbf{a} Kramers Weyl materials with nonmagnetic chiral point group symmetries: The Weyl points, represented by the red and blue dots, are pinned at time-reversal-invariant momenta (TRIMs).
		\textbf{b} Kramers nodal line materials with nonmagnetic achiral point group symmetries: There are doubly degenerate lines (blue lines) joining the TRIMs. These nodal lines are called Kramers nodal lines.
		\textbf{c} Heesch Weyl antiferromagnets (AFMs) with inadmissible chiral point group symmetries: Two-dimensional irreducible corepresentations of magnetic little groups enforce Heesch Weyl points (red and blue dots) at high symmetry momenta.
		\textbf{d} Heesch nodal line antiferromagnets with inadmissible achiral point group symmetries: There are doubly degenerate lines (blue lines) joining high symmetry points with two-dimensional irreducible corepresentations. These doubly degenerate lines are called Heesch nodal lines.
		The insets in \textbf{c} and \textbf{d} schematically show representative chiral and achiral AFM orders.}
	\label{fig_four_types}
\end{figure}
 
 Conventional Weyl points and nodal lines usually appear at generic momenta or along generic paths, although their existence may be inferred from symmetry-based indicators \cite{Po2017, PKruthoff2017, Watanabe2018, Elcoro2021}. 
 By contrast, little group analysis at high symmetry momenta can identify symmetry-pinned Weyl points and nodal lines emanating from them \cite{Chang2018, Xie2021, gao2023}.
 In nonmagnetic chiral crystals with spin-orbit coupling, time-reversal symmetry enforces Kramers degeneracies at time-reversal-invariant momenta. 
 The absence of achiral symmetries allows these degeneracies to carry nonzero topological charges and become Weyl points pinned at those momenta, known as Kramers Weyl fermions~\cite{Chang2018} (Fig.~\ref{fig_four_types}a).  
 Restoring achiral symmetries such as mirror or rotoinversion can connect and annihilate Kramers Weyl fermions, resulting in Kramers nodal lines joining time-reversal-invariant momenta~\cite{Xie2021} (Fig.~\ref{fig_four_types}b).
 Kramers nodal lines are protected by time-reversal symmetry and achiral point-group symmetries, with experimental evidence reported in several recent studies~\cite{Zhang2023, Sarkar2023, Kurumaji2025, Zhang2025, Gabriele2025, Sarkar2026}.
 
 Magnetic order further enriches the topological properties by breaking time-reversal symmetry and introducing magnetic crystalline symmetries.
 In inadmissible chiral antiferromagnets, two-dimensional double-valued irreducible corepresentations (IR coreps) of magnetic little groups enforce the presence of Weyl fermions at high symmetry momenta, known as Heesch Weyl (HW) points \cite{gao2023} (Fig.~\ref{fig_four_types}c).
 Here, inadmissibility means that the corresponding magnetic point group (MPG) is incompatible with ferromagnetic order and forbids a nonzero net magnetization.
 Moreover, the HW points have to be removed from the high symmetry points with double-valued IR coreps if the MPG becomes achiral (when additional symmetries such as mirror symmetry are restored, for example). 
 However, as the original HW points are pinned at high symmetry momenta by two-dimensional IR coreps of the magnetic little groups, adding additional achiral symmetries cannot remove the double degeneracy of the original HW points.
 Therefore, it is puzzling how the HW points can possibly be removed when the material evolves from a chiral inadmissible antiferromagnet to an achiral inadmissible antiferromagnet. 
 
 In this work, we resolve the above puzzle by showing that when a chiral inadmissible antiferromagnet evolves into an achiral inadmissible antiferromagnet, doubly degenerate lines emerge to connect pairs of HW points such that the HW points are removed. 
 We call the doubly degenerate lines the Heesch nodal lines (HNLs), and the materials hosting them the Heesch nodal line antiferromagnets. 
 A schematic picture of the HNLs is shown in Fig.~\ref{fig_four_types}d. 
 We emphasize that the high symmetry point degeneracies connected by HNLs are protected by two-dimensional IR coreps of the corresponding magnetic little groups, while the HNLs are enforced by the additional achiral symmetries, as illustrated in Fig.~\ref{fig_symmetry}. 
 
 We identify all the relevant inadmissible achiral MPGs and determine the allowed HNL directions by constructing the low-energy $\bm{k}\cdot\bm{p}$ Hamiltonians (Tables~\ref{tb_kp_colorless} and \ref{tb_kp_bnw}). 
 As an explicit illustration, we construct a tight-binding model to show that the HNLs emerge and connect the original HW points by restoring achiral symmetries, as shown in Fig.~\ref{fig_HAHC}a. 
 Guided by group-theoretical analysis, we identify candidate antiferromagnetic materials hosting HNLs, as summarized in Table~\ref{tb_materials}.
 We then confirm the predicted HNLs by first-principles calculations for MnTe, CrSb, and Mn$_3$GaN, establishing these materials as promising platforms for observing and engineering HNLs (Fig.~\ref{fig_materials}).
 For antiferromagnetic crystals hosting HNLs, symmetry constraints make higher-order anomalous Hall effects the leading allowed Hall responses (Fig.~\ref{fig_HAHC}).
 In most inadmissible achiral MPGs, this leading response is the third-order nonlinear Hall effect, distinguishing HNL antiferromagnets from Kramers nodal line systems.
 Moreover, perturbations such as strain can lower an inadmissible MPG to an admissible subgroup and split the HNLs.
 The low-energy physics of split HNLs is described by a collection of two-dimensional massive Dirac Hamiltonians.
 Importantly, broken time-reversal symmetry permits nonzero Chern numbers of two-dimensional planes in the Brillouin zone and gives rise to chiral edge states on the surfaces of the antiferromagnets (Fig.~\ref{fig_finite_chirality}).
 This provides a route to engineering anomalous Hall effects in antiferromagnetic crystals.

\section{Results}
\subsection{Symmetry protected Heesch nodal lines} 
\label{sec:symm}
\begin{figure}[h]
	\centering
	\includegraphics[width=1\columnwidth]{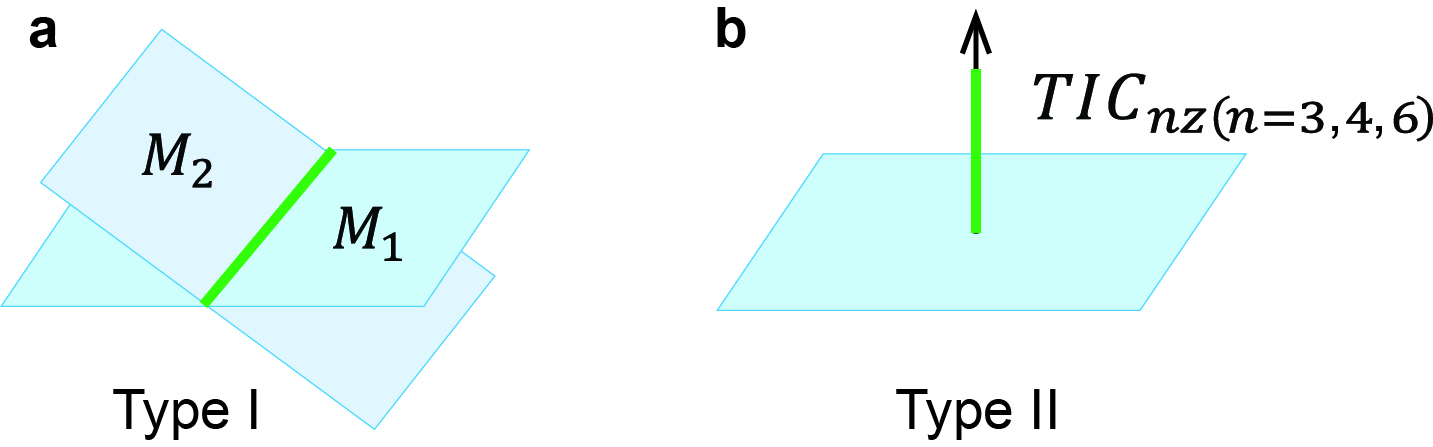}
    \caption{\textbf{Two types of Heesch nodal lines.}
    \textbf{a} A type I HNL (green line) lies at the intersection of two mirror planes, $M_1$ and $M_2$.
    \textbf{b} A type II HNL (green line) lies along the rotation axis and is enforced by the symmetry operation ${T}IC_{nz}$, with $n=3,4,6$.} 
	\label{fig_symmetry}
\end{figure}

In this section, we establish the group-theoretical foundation for the emergence of HNLs in achiral antiferromagnets.  
Inadmissible MPGs lacking both time-reversal symmetry and parity-time symmetry can host two-dimensional IR coreps, which enforce robust twofold degeneracies at high symmetry momenta without relying on Kramers degeneracy.
Near these degenerate momenta, achiral symmetry constraints on the $\bm{k}\cdot\bm{p}$ Hamiltonian can forbid band splitting terms along high symmetry lines, thereby extending the point degeneracy into an HNL.
Depending on the underlying symmetry mechanism, HNLs can be classified into type I and type II, as illustrated in Fig.~\ref{fig_symmetry}.

\textbf{Inadmissibility-enforced double degeneracies at high symmetry momenta.} 
We focus on the 122 MPGs rather than magnetic space groups (MSGs) to avoid additional complications from nonsymmorphic operations. 
Among them, 31 MPGs are admissible and host only one-dimensional IR coreps.  
From the remaining 91 inadmissible MPGs, we further exclude those preserving time-reversal symmetry or parity-time symmetry: the former correspond to nonmagnetic grey groups hosting Kramers Weyl fermions and Kramers nodal lines \cite{Chang2018,Xie2021}, while the latter lead to globally doubly degenerate bands. 
The remaining 38 relevant inadmissible MPGs (11 chiral MPGs and 27 achiral MPGs) can be classified into colorless MPGs, which contain only unitary elements $\hat g$, and black-and-white MPGs, which also contain antiunitary elements $T\hat g$. 
Although these groups lack Kramers degeneracy due to broken time-reversal symmetry, their inadmissibility guarantees the existence of at least one two-dimensional IR corep, which enforces a robust twofold degeneracy at the corresponding high symmetry momenta.

The origin of two-dimensional IR coreps is different for colorless and black-and-white MPGs. 
For colorless MPGs, all symmetry operations are unitary. 
A two-dimensional IR corep appears when the corepresentation matrices of the generators cannot be simultaneously reduced to one-dimensional blocks by any basis transformation. 
The situation is different for antiunitary symmetry operations because they involve complex conjugation and cannot be treated by ordinary simultaneous diagonalization. 
In this case, even a single antiunitary generator can enforce a two-dimensional IR corep. For example, in the MPG $\bar{4}^{\prime}$, the single antiunitary generator $T I C_{4z}$ enforces two-dimensional IR coreps.
This irreducibility reflects the inadmissibility of MPGs and gives rise to protected twofold degeneracies at high symmetry momenta.
A more rigorous derivation is provided in Supplementary Note 2.

\begin{table*}[!t]
 	\renewcommand{\arraystretch}{1.2}
 	\centering
 	\caption{\textbf{Heesch nodal lines in candidate materials.} 
 	The definitions of high symmetry momenta follow the conventions given in the Bilbao Crystallographic Server \cite{Bilbao}. 
 	Representative achiral antiferromagnetic materials hosting Heesch nodal lines (HNLs) are identified with the assistance of MAGNDATA \cite{MAGNDATA_I,MAGNDATA_II}.
 	The corresponding magnetic space groups (MSGs) and Belov--Neronova--Smirnova (BNS) numbers are also listed.
 	Parentheses in the HNLs column indicate either Heesch Weyl fermions (HWFs) at the indicated high symmetry momenta or HNLs protected by a different magnetic point group (MPG).}
 	\label{tb_materials}
 		\begin{tabular}[t]{l l l l l}
 			\hline\hline
 			Type &  MSG (BNS no.) & MPG & HNLs & Materials 
 			\\ 
 			\hline
 			Colorless &
 			$\rm Pmc2_1 (26.66)$  & $mm2$
 			& $\rm \Gamma\text{-}Z, Y\text{-}T, S\text{-}R, X\text{-}U$ 
 			& $\rm FeSb_2O_4$, $\rm FePbBiO_4$
 			\\&
 			$\rm  Pnma (62.441) $ & $mmm$  
 			& $\rm \Gamma\text{-}Z, \Gamma\text{-}Y, \Gamma \text{-}X, X\text{-}S,  X\text{-}U, U\text{-}Z, U\text{-}R, R\text{-}T $
 			& $\rm Mn_2GeO_4, Co_2SiO_4$
 			
 			\\&
 			$\rm Cmcm (63.457)$ & $mmm$  
 			& $\rm \Gamma\text{-}Z, \Gamma\text{-}Y, \Gamma \text{-}X, Y\text{-}T, Y\text{-}X_1, Z\text{-}T$
 			& $\rm MnTe, NiCrO_4$
 			\\ & 
 			$\rm P\bar{4}2_1m(113.267)$ & $\bar{4}2m$ 
 			& $\rm \Gamma\text{-}Z,M\text{-}A$ 
 			& $\rm Ba_2MnSi_2O_7$
 			\\ & 
 			$\rm I4_1/amd (141.551)$ & $4/mmm$ 
 			& $\rm \Gamma\text{-}Z, \Gamma\text{-}M, \Gamma\text{-}X$ &$\rm CdYb_2S_4$, $\rm CdYb_2Se_4 $ 
 			\\&
 			$\rm P31m (157.53)$ & $3m$ 
 			& $\rm \Gamma\text{-}A,H\text{-}K, HA\text{-}KA$ 
 			& $\rm Ba_3MnNb_2O_9$
 			\\ &
 			$\rm R3c (161.69)$ & $3m$ 
 			&$\rm \Gamma\text{-}T$  
 			& $\rm PbNiO_3$,  $\rm Ca_3Co_{2-x}Mn_xO_6$
 			\\&
 			$\rm R\bar{3}m(166.97)$  &$\bar{3}m$ 
 			& $\rm \Gamma\text{-}T$ 
 			& $\rm Mn_3GaN$, $\rm Mn_3ZnN$   
 			\\ &
 			$\rm R\bar{3}c (167.103)$ &$\bar{3}m$ &$\rm \Gamma\text{-}T$ 
 			&$\rm FeCO_3$, $\rm CoF_3$, $\rm LaCrO_3$
 			\\&
 			$\rm P6_3cm (185.197)$ &$6mm$  
 			& $\rm \Gamma\text{-}A$, ${\rm K\text{-}H} (3m) $,  ${\rm M\text{-}L} (mm2)$ 
 			&$\rm YMnO_3$, $\rm HoMnO_3$
 			\\&
 			$\rm P\bar{6}2m (189.221)$ & $\bar{6}m2$  
 			&$\rm\Gamma\text{-}A, \Gamma\text{-}K, A\text{-}H$
 			& $\rm Ba_3CoSb_2O_9$
 			\\ & 
 			$\rm Pa\bar{3} (205.33)$ &$m\bar{3}$ 
 			&${\rm\Gamma\text{-}X, X\text{-}M}(mmm)$ 
 			& $\rm MnTe_2 $, $\rm NiS_2$
 			\\  
 			\hline
 			Black and white& $\rm I\bar{4}'(82.41)$ & $\bar{4}^\prime$ 
 			& $\rm\Gamma\text{-}M$ 
 			& $\rm CsCoF_4$
 			\\& 
 			$\rm I\bar{4}'2d' (122.336)$ & $\bar{4}^{\prime} 2m^{\prime}$ 
 			& $\rm\Gamma\text{-}Z$ 
 			& $\rm Ce_4Sb_3$
 			\\  &
 			$\rm I4'/mm'm (139.534)$ & $4^{\prime}/mm^{\prime}m$ 
 			&${\rm \Gamma\text{-}M, \Gamma\text{-}X, \Gamma\text{-}Z,\rm X\text{-}P} (mmm)$ 
 			& $\rm La_2O_3Mn_2Se_2$
 			\\&
 			$\rm I4_1'/am'd (141.554)$ & $4^{\prime}/mm^{\prime}m$ & $\rm \Gamma\text{-}M, \Gamma\text{-}X, \Gamma\text{-}Z$ & $\rm Er_2Ru_2O_7$
 			\\&
 			$\rm P\bar{6}' (174.135)$ &$\bar{6}^{\prime}$ 
 			&$\rm \Gamma\text{-}A$  
 			& $\rm Tb_{14}Ag_{51}$, 
 			$\rm Cu_{0.82}Mn_{1.18}As$
 			\\  &
 			$\rm P6_3'm'c (186.205)$ & $6^{\prime}mm^{\prime}$ 
 			& $\rm \Gamma\text{-}A, H\text{-}K, HA\text{-}KA$ 
 			& $\rm Fe_2Mo_3O_8$, $\rm Co_2Mo_3O_8 $
 			\\  &
 			$\rm P\bar{6}'2m' (189.224)$ & $\bar{6}^{\prime} m^{\prime} 2$ 
 			&$\rm \Gamma\text{-}A (\rm HWFs: H,K,HA,KA )$  
 			&$\rm RbFeCl_3$, $\rm UNiGa$, $\rm TmAgGe$  
 			\\ & 
 			$\rm P\bar{6}'2'm (189.223)$ &$\bar{6}^{\prime} m2^{\prime}$
 			&${ \rm \Gamma\text{-}A, \rm H\text{-}K}(3m),{\rm HA\text{-}KA}(3m)$  
 			& $\rm ThMn_2$, $\rm CsFeCl_3$
 			\\  &
 			$\rm P6_3'/m'm'c (194.268)$ & $6^{\prime}/m^{\prime}mm^{\prime}$
 			&$\rm \Gamma\text{-}A$
 			& $\rm CrSb$, $\rm CrNb_4S_8$
 			\\ & 
 			$\rm Fd\bar{3}m' (227.131)$ & $m\bar{3}m^{\prime}$ 
 			& $\rm \Gamma\text{-}X$ & $\rm Cd_2Os_2O_7$, $\rm Yb_2Ir_2O_7$, $\rm Nd_2Ir_2O_7$
 			\\ 
 			\hline\hline
 		\end{tabular}
 \end{table*}
 
 \textbf{Two types of HNLs enforced by achiral symmetry.}    
 We now show how achiral symmetries extend the point degeneracies protected by inadmissibility into two types of HNLs. The generic two-band Hamiltonian near a degenerate momentum $\bm{k}_0$ can be written as
 \begin{equation}
 	H(\bm{k}) = f_0(\bm{k})\sigma_0+ \bm{f}(\bm{k})\cdot \bm{\sigma},
 \end{equation}
 where $\bm{k}$ is measured relative to $\bm{k}_0$. 
 We use the helical basis $\bm{k}= [k_+,k_-,k_z]^T$, $\bm{f}= [f_+,f_-,f_z]^T$, and $\bm{\sigma}= [\sigma_+,\sigma_-,\sigma_z]^T$, with $k_\pm = k_x \pm i k_y$ and $\sigma_\pm = \sigma_x \pm i\sigma_y$. 
 $\bm{f}(\bm{k})\cdot\bm{\sigma}$ represents the spin-dependent part of the Hamiltonian arising from spin-orbit coupling or antiferromagnetic order. 
 A symmetry operation $\hat{g}\in G_{\bm{k}_0}$ imposes the constraint
 \begin{equation}\label{eq_fk}
 	\bm{f}(\bm{k})  = \det(\hat{g})\hat{g}^T\bm{f}(\hat{g}\bm{k})  .
 \end{equation}    
 Equation~\eqref{eq_fk} shows how symmetries constrain the allowed spin polarization near $\bm{k}_0$, thereby determining whether nodal lines can emanate from the degenerate momentum. 
 A detailed derivation for different symmetry operations is provided in Supplementary Note 3. 

\begin{figure*}[htbp]
	\centering
	\includegraphics[width=1\linewidth]{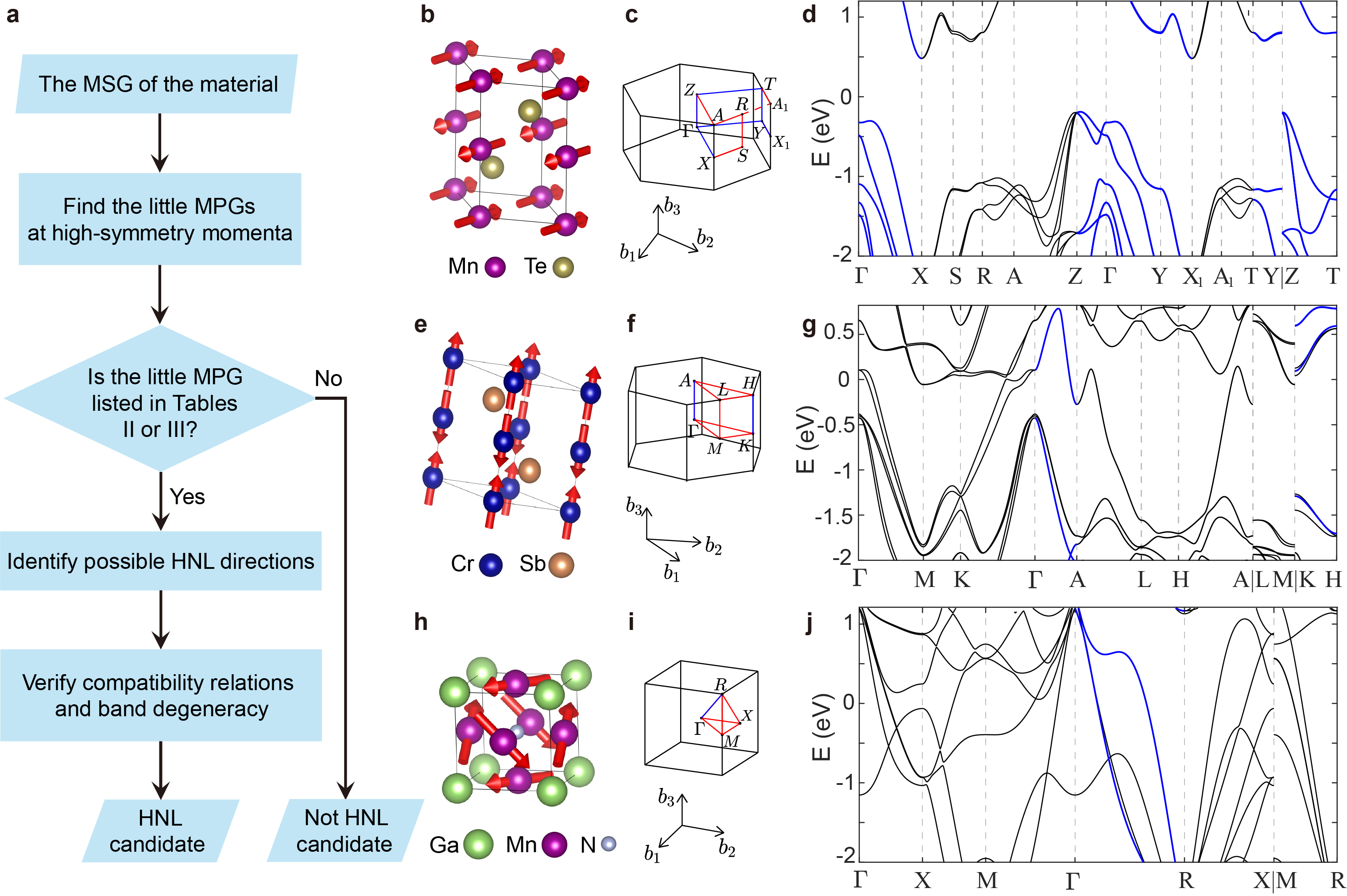}
	\caption{\textbf{Candidate antiferromagnetic materials with Heesch nodal lines.} 
	\textbf{a} Flowchart for the systematic identification of Heesch nodal lines (HNLs) from a given magnetic space group (MSG). MPG denotes magnetic point group. 
	\textbf{b--j} The crystal structures, first Brillouin zones, and electronic band structures for three example candidate materials: \textbf{b--d} collinear antiferromagnet MnTe, \textbf{e--g} collinear antiferromagnet CrSb, and \textbf{h--j} noncollinear antiferromagnet Mn$_3$GaN. 
	In the crystal structures \textbf{b,e,h}, red arrows depict atomic magnetic moments. 
	In the Brillouin zones \textbf{c,f,i}, red lines are high symmetry paths and blue lines denote the predicted HNLs. 
	In the band structures \textbf{d,g,j}, HNLs are highlighted in blue.}
	\label{fig_materials}
\end{figure*}

 In achiral MPGs, mirror ($M$) and antiunitary rotoinversion symmetries ($TIC_{nz}$) can give rise to two types of HNLs, classified as type I and type II, as illustrated in Fig.~\ref{fig_symmetry}.
 Type I HNLs emerge along the intersection of two distinct mirror-invariant planes. 
 For example, in the presence of the mirror symmetry $M_z$, Equation~\eqref{eq_fk} forces $f_\pm=0$ on the $k_z=0$ plane, leaving only the out-of-plane spin polarization $f_z$. 
 When two such planes intersect, their symmetry constraints can jointly force all components of $\bm f(\bm k)$ to vanish along the intersection line, forming a symmetry-enforced nodal line emanating from the degenerate momentum (Fig.~\ref{fig_symmetry}a).  
 Type II HNLs originate from a different mechanism. 
 Under an antiunitary symmetry operation $TIC_{nz}$ with $n=3,4,6$, Equation~\eqref{eq_fk} forces all components $f_\pm$ and $f_z$ to vanish along the rotation axis $k_x=k_y=0$. 
 This directly enforces a degenerate line along the $\hat{z}$ axis, forming a type II HNL, as depicted in Fig.~\ref{fig_symmetry}b. 
 In general, all HNLs in colorless MPGs belong to type I (Table~\ref{tb_kp_colorless}). 
 In contrast, black-and-white MPGs can host both type I and type II HNLs (Table~\ref{tb_kp_bnw}).
 
 To determine the directions of HNLs emanating from high symmetry momenta, we construct the $\bm{k}\cdot\bm{p}$ Hamiltonians for two-dimensional IR coreps of all relevant inadmissible achiral MPGs, as summarized in Tables~\ref{tb_kp_colorless} and \ref{tb_kp_bnw}. 
 The resulting nodal lines generally lie along high symmetry lines, consistent with the arguments above.

\subsection{Candidate materials hosting Heesch nodal lines}  
To identify candidate materials hosting HNLs, we follow the workflow summarized in Fig.~\ref{fig_materials}a. 
Starting from the MSG of a material, we determine the little MPG at each high symmetry momentum and compare it with the MPGs hosting HNLs in Tables~\ref{tb_kp_colorless} and \ref{tb_kp_bnw}. 
The corresponding table entries give the possible HNL directions and effective Hamiltonians.
In our convention, the principal symmetry axis is chosen as the $\hat{z}$ direction, whereas the corresponding axes in a candidate material must be identified according to the actual symmetry operations.
Nonsymmorphic operations can introduce extra phase factors and modify the compatibility relations especially at the Brillouin-zone boundaries, potentially altering or lifting the HNLs
(see the detailed discussion in Supplementary Note 5).
Finally, we verify the candidates by checking compatibility relations and band degeneracies along the predicted high symmetry lines \cite{book_group,Bilbao}. 
Guided by this symmetry-based protocol, we search the magnetic materials database \cite{MAGNDATA_I,MAGNDATA_II} and identify several inadmissible achiral antiferromagnets as candidate materials for hosting HNLs along high symmetry directions. 
The resulting candidates, covering both symmorphic and nonsymmorphic crystals, are summarized in Table~\ref{tb_materials}.  
In the following, we take three representative materials as examples and confirm the predicted HNLs by first-principles calculations, as shown in Fig.~\ref{fig_materials}b--j.

MnTe provides a representative example of a collinear antiferromagnet, as shown in Fig.~\ref{fig_materials}b. 
Its paramagnetic parent phase crystallizes in the space group $\rm P6_{3}/mmc$. 
Below the N\'eel temperature ($T_{\rm N}=323~\mathrm{K}$), MnTe develops collinear antiferromagnetic order described by the MSG $\rm Cmcm$ and the MPG $mmm$. 
At each of the high symmetry momenta $\Gamma$, $\rm T$, $\rm Y$, and $\rm Z$, the little MPG is $mmm$. 
The three mirror-invariant planes of this group give rise to type I HNLs along their intersection lines, leading to the twofold band degeneracies shown in Fig.~\ref{fig_materials}d.
Notably, nonsymmorphic phase factors modify the compatibility relations on the $k_z=\pi/c$ plane, so that the HNL degeneracies are not maintained along the $\rm Z\text{-}A$ and $\rm T\text{-}A_1$ paths (see Supplementary Note 5). 
 
CrSb provides another representative example of a collinear antiferromagnet, as shown in Fig.~\ref{fig_materials}e. It has recently been identified as a prototypical altermagnetic material \cite{PhysRevLett.133.206401,Reimers2024,Yang2025}.
The HNLs in this system can be viewed as remnants of the nodal planes expected in ideal altermagnets after the inclusion of spin-orbit coupling \cite{PhysRevB.109.024404}.
In the paramagnetic phase, CrSb shares the same space group as MnTe. 
Below its N\'eel temperature ($T_{\rm N}=600~\mathrm{K}$), it develops collinear antiferromagnetic order along the $\hat{z}$ axis, reducing the symmetry to MSG $\rm P6_3^\prime/m^\prime m^\prime c$ with the black-and-white MPG $6^\prime/m^\prime mm^\prime$. 
 The high symmetry momenta $\Gamma$ and $\rm A$ inherit the full MPG $6^\prime/m^\prime mm^\prime$, which admits both one- and two-dimensional IR coreps. 
 The latter lead to HNLs along the $\hat{z}$ axis, corresponding to the $\Gamma\text{-}\rm A$ segment shown in Fig.~\ref{fig_materials}f,g. 
 In addition, the MSG operation $\{TC_{6z}|00\tfrac{1}{2}\}$ contains a half translation along the $\hat{z}$ direction, enforcing double degeneracy in the boundary plane of the Brillouin zone $k_z=\pi/c$. 
 Details of the nonsymmorphic effects are given in Supplementary Note 5.
 The high symmetry momenta $\rm K$ and $\rm H$, characterized by the little MPG $6^\prime mm^\prime$, also support HNLs along the $\rm K\text{-}H$ direction (Fig.~\ref{fig_materials}f,g). 

Unlike MnTe and CrSb, Mn$_3$GaN exhibits highly noncollinear magnetic order, as shown in Fig.~\ref{fig_materials}h.
Its high-temperature paramagnetic phase adopts a cubic structure with space group $\rm Pm\bar{3}m$. 
Below its N\'eel temperature ($T_N=298~\mathrm{K}$), the magnetic moments of the Mn atoms lie within the $(111)$ plane and point along the face-diagonal directions.  
This magnetic order is described by the symmorphic MSG $\rm R\bar{3}m$ and the colorless MPG $\bar{3}m$.  
Within each $(111)$ Mn layer, the Mn moments form a frustrated antiferromagnetic kagome pattern.  
In the convention of Table~\ref{tb_materials}, the HNL is listed along $\Gamma\text{-}\rm T$, where the principal symmetry axis is chosen as $\hat z$. 
For Mn$_3$GaN, this axis corresponds to the threefold rotation axis along the $[111]$ direction, namely the $\Gamma\text{-}\rm R$ path in Fig.~\ref{fig_materials}i,j. 
The high symmetry momenta $\Gamma$ and $\rm R$ inherit the full MPG $\bar{3}m$, which admits both one- and two-dimensional IR coreps. 
Accordingly, both split and degenerate bands are present along the $\Gamma\text{-}\rm R$ path, as shown in Fig.~\ref{fig_materials}j.

\subsection{Nonlinear anomalous Hall effect in Heesch nodal line antiferromagnets}
\label{sec:HAHE}
The linear anomalous Hall effect is permitted only in the 31 admissible MPGs, while it is forbidden in inadmissible antiferromagnetic crystals hosting HNLs \cite{Smejkal2022}.
This prohibition stems from the pseudovector nature of the Berry curvature. 
In inadmissible antiferromagnetic MPGs, the symmetry operations responsible for inadmissibility force the Brillouin-zone integral of the Berry curvature to vanish, leading to zero linear anomalous Hall conductivity.
  
Although the linear anomalous Hall effect is absent, nonlinear anomalous Hall responses induced by Berry curvature multipoles can become the leading allowed Hall responses \cite{PhysRevLett.115.216806,PhysRevB.107.115142}.
Through a detailed symmetry analysis (see Methods and Supplementary Note 4), we determine the allowed Berry curvature dipole and quadrupole components for all MPGs hosting HNLs, as listed in Supplementary Table~1.
Only 10 HNL MPGs admit a nonzero Berry curvature dipole $D_{\alpha\beta}$ and can therefore host a second-order nonlinear Hall effect.
This restricted second-order response differs markedly from that in the MPGs of HW antiferromagnets, because achiral symmetries strongly constrain $D_{\alpha\beta}$ \cite{gao2023}.
In contrast, a nonzero Berry curvature quadrupole and the associated third-order nonlinear anomalous Hall effect are permitted in most HNL MPGs.
This third-order response distinguishes HNL antiferromagnets from Kramers Weyl and Kramers nodal line systems, where time-reversal symmetry is preserved and the Berry curvature quadrupole is forbidden \cite{PhysRevB.107.115142,Chang2018,Xie2021}. 

\begin{figure*}[!t]
	\centering
	\includegraphics[width=1\linewidth]{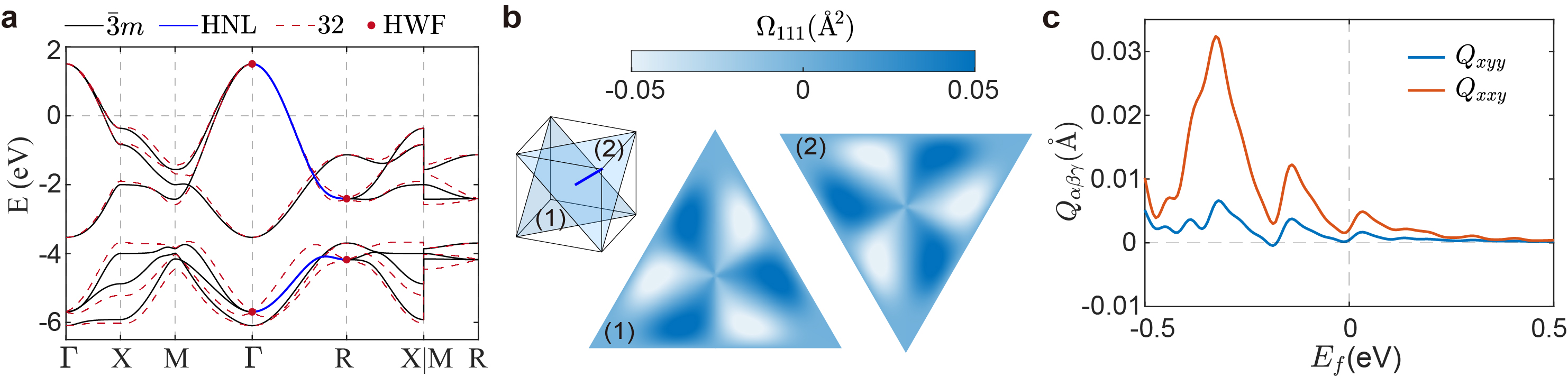}
	\caption{\textbf{Higher-order nonlinear anomalous Hall effect in Heesch nodal line antiferromagnets.} 
		\textbf{a} Band structures calculated from the tight-binding Hamiltonian for magnetic point groups $\bar{3}m$ and $32$, with Heesch nodal lines (HNLs) highlighted by blue lines and Heesch Weyl fermions (HWFs) marked by dark red dots. 
		\textbf{b} Berry curvature component along the $[111]$ direction, $\Omega_{111}$, in two momentum planes perpendicular to $[111]$. The left cube shows the Brillouin zone, with the two representative momentum planes highlighted in blue. 
		\textbf{c} Chemical-potential dependence of the Berry curvature quadrupole components $Q_{xyy}$ and $Q_{xxy}$. 
		Parameters: $a = 3.898~$\AA, $\epsilon_0 = 3~\rm{eV}$, $t_1=0.5~\rm{eV}$, $t_2 = 0.6~\rm{eV}$, $J = 1~\rm{eV}$, $t_{\text{SOC},1}=0.1~\rm{eV}$, $\phi_\text{SOC} = \pi/3$, $t_{\text{SOC},2}=0.02~\rm{eV}$, $\delta t_1 = 0.3t_1$, $\delta t_2 = 0.3t_2$, $k_BT= 0.02~\rm{eV}$.}
	\label{fig_HAHC}
\end{figure*} 

To illustrate the nonlinear anomalous Hall response in HNL antiferromagnets, we take Mn$_3$GaN as a concrete example.
We construct a double-exchange $s$-$d$ model describing itinerant $s$ electrons coupled to localized Mn $d$ magnetic moments \cite{PhysRevB.62.R6065, Chen2014}. 
The MPG $\bar{3}m$ is generated by a threefold rotation $C_{3,111}$, a twofold rotation $C_{2,1\bar{1}0}$, and spatial inversion $I$.
We include a spin-orbit coupling term consistent with the MPG $\bar{3}m$ [Eq.~\eqref{eq_HSOC} in Methods], which generates locally nonzero Berry curvature.
The calculated band structure exhibits HNLs, highlighted by blue lines in Fig.~\ref{fig_HAHC}a. 
Figure~\ref{fig_HAHC}b shows the Berry curvature component along the $[111]$ direction, $\Omega_{111}$, in two representative momentum slices for the top band.
The combined $C_{3,111}$ and $C_{2,1\bar{1}0}$ symmetries forbid a net axial vector, leading to a vanishing linear anomalous Hall conductivity.
Inversion symmetry further forbids a Berry curvature dipole.
Nevertheless, the symmetry allows a nonzero Berry curvature quadrupole, leading to a third-order anomalous Hall response.  
Figure~\ref{fig_HAHC}c shows the calculated Berry curvature quadrupole components $Q_{xyy}$ and $Q_{xxy}$ as functions of the chemical potential.

Breaking the achiral symmetries of $\bar{3}m$ reduces the MPG to the chiral MPG $32$, which hosts HW points. 
As shown in Fig.~\ref{fig_HAHC}a, the HNL along $\Gamma\text{-}\rm R$ (blue lines) in $\bar{3}m$ is lifted except at $\Gamma$ and $\rm R$, leaving two HW points pinned at these high symmetry momenta (dark red dots). 
In the chiral MPG $32$, the allowed Berry curvature dipole leads to a second-order nonlinear Hall response, in contrast to the leading third-order response in the $\bar{3}m$ HNL case.

\begin{figure*}[htbp]
	\centering
	\includegraphics[width=1\linewidth]{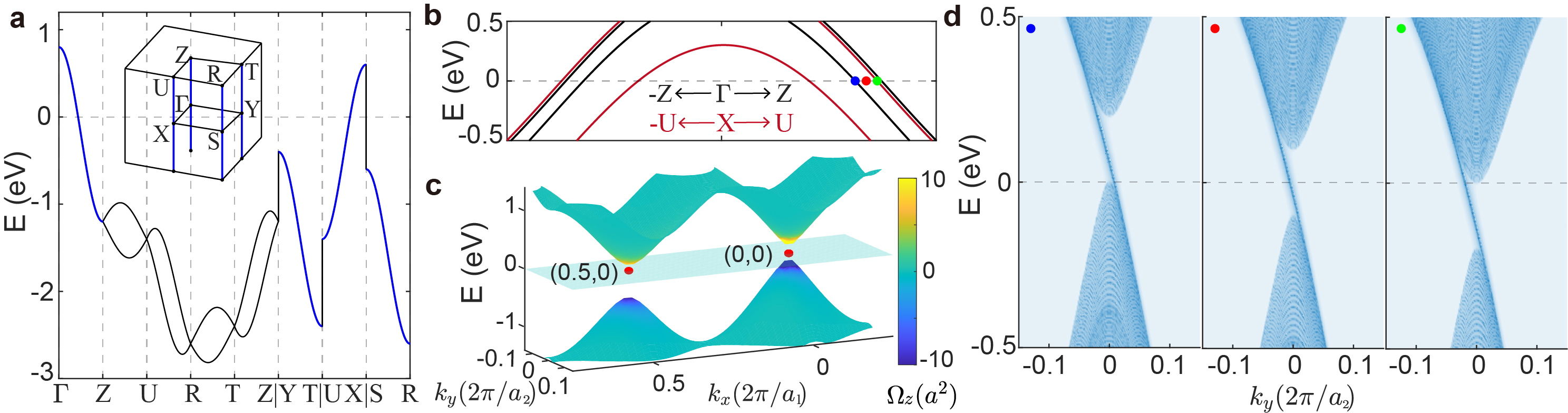}
	\caption{\textbf{Chiral edge states induced by split Heesch nodal lines.} 
		\textbf{a} Band structure from the tight-binding Hamiltonian, with blue lines highlighting the Heesch nodal lines. 
		\textbf{b} Split nodal lines near the Fermi energy along $\rm -Z\text{-}\Gamma\text{-}Z$ and $\rm -U\text{-}X\text{-}U$. Colored dots mark representative $k_z$ momenta. 
		\textbf{c} Energy dispersion $E(k_x,k_y)$ and Berry curvature $\Omega_z(k_x,k_y)$ at the $k_z$ momentum marked by the red dot in \textbf{b}. 
		\textbf{d} The local density of states on the left edge, calculated with open boundary conditions along the $\hat{x}$ direction, for the $k_z$ momenta marked in \textbf{b}. 
		Parameters: $\epsilon_0 = -0.8~\mathrm{eV}$, $t_z=0.5~\mathrm{eV}$, $t_x = 0.05~\mathrm{eV}$, $t_y = 0.3~\mathrm{eV}$, $\alpha_1 =0.4~\mathrm{eV}$, $\alpha_2 = 0.3~\mathrm{eV}$, $\alpha_3 = \alpha_4 = \alpha_5 = 1~\mathrm{eV}$, $\beta_1 = \beta_x =0.1~\mathrm{eV}$, $\beta_2 = 0.02~\mathrm{eV}$, and $a_1=a_2=a_3=1$.}
	\label{fig_finite_chirality}
\end{figure*}

\subsection{Chiral edge states induced by the splitting of Heesch nodal lines}\label{sec:AHE_NLsplitting}
Most inadmissible achiral MPGs contain at least one inadmissible chiral subgroup, as indicated in Tables~\ref{tb_kp_colorless} and~\ref{tb_kp_bnw}. 
This implies that breaking the corresponding achiral symmetries can lift the HNL along the high symmetry line, leaving HW points at the connected high symmetry momenta protected by two-dimensional IR coreps.  
Further breaking the symmetries protecting these two-dimensional IR coreps removes the remaining HW points and reduces the MPG to an admissible subgroup, leading to complete lifting of the line degeneracy.
Near the Fermi-level crossings of an HNL, the low-energy physics is described by two-dimensional massless Dirac cones for planes intercepting the HNL.
Splitting the HNL opens gaps in these Dirac cones. 
This gapping generates Berry curvature near the massive Dirac cones and can yield a nonzero anomalous Hall conductivity in these antiferromagnetic crystals.

We illustrate this physics using the MPG $mm2$ as a concrete example.
The low-energy Hamiltonian consists of three terms 
\begin{equation}
	H =	H_0+ H_{\rm SOC}+	H_{\rm  M}  .
\end{equation}
Here, $H_0 = (-\epsilon_0 -t_x^2 k_x^2 -t_y^2k_y^2 -t_z^2 k_z^2)\sigma_0$ represents the spin-independent dispersion, $H_{\rm SOC} = \alpha_1 k_y \sigma_x +\alpha_2 k_x \sigma_y $ describes the spin-orbit coupling, and $H_{\rm  M} = \alpha_3 k_y k_z \sigma_x +\alpha_4 k_x k_z \sigma_y + \alpha_5 k_x k_y \sigma_z$ is the second-order spin-dependent term arising from antiferromagnetic order. 
We implement this low-energy Hamiltonian in a tight-binding model on a primitive orthorhombic lattice. 
The relevant high symmetry points retain the little MPG $mm2$, with the principal axis oriented along the $\hat z$ direction. 
Consequently, four HNLs appear along high symmetry lines parallel to the $\hat z$ direction, as highlighted by the blue lines in Fig.~\ref{fig_finite_chirality}a. 
In particular, the Fermi energy crosses two of the HNLs along the $\Gamma\text{-}\rm Z$ and $\rm X\text{-}U$ paths.

The application of in-plane strain breaks the mirror symmetries of the crystal, reducing the MPG from $mm2$ to $2$. 
The corresponding low-energy perturbation is given by 
$\delta H_2 =(\beta_1 +\beta_2 k_z - \beta_x k_x^2)\sigma_z$.  
Figure~\ref{fig_finite_chirality}b shows the resulting split HNLs near the Fermi energy, where the three dots mark representative $k_z$ momenta. 
At the momentum marked by the red dot, the Fermi energy lies inside the gap opened by the splitting of the nodal lines. 
Figure~\ref{fig_finite_chirality}c shows the energy dispersion $E(k_x,k_y)$ at this fixed $k_z$ momentum. 
The splitting of the nodal lines gaps out the low-energy massless Dirac cones. 
Each gapped Dirac point contributes $-1/2$ to the Chern number of the valence band, yielding a total Chern number of $-1$ for this two-dimensional momentum slice, as shown in Fig.~\ref{fig_finite_chirality}c.
In the middle panel of Fig.~\ref{fig_finite_chirality}d, we impose open-boundary conditions along the \(\hat{x}\) direction and compute the local density of states on the left edge using the surface Green's function method \cite{Sancho_1984, Sancho_1985}. The result shows that the Fermi energy intersects the chiral edge states.  
For other $k_z$ momenta, such as those marked by the blue and green dots in Fig.~\ref{fig_finite_chirality}b,d, bulk conducting channels can also coexist with the chiral edge states. 

\section{Discussion and Conclusion}  
In summary, we have systematically identified HNLs in inadmissible achiral antiferromagnets based on MPG analysis and provided a series of candidate materials hosting HNLs. 
We conclude that all inadmissible antiferromagnets without parity-time symmetry are topological: inadmissible chiral antiferromagnets host HW points, whereas inadmissible achiral antiferromagnets host HNLs.
As HNLs lie along high symmetry paths in the Brillouin zone, they provide promising targets for observation using angle-resolved photoemission spectroscopy, analogous to previous experimental studies of Kramers nodal lines \cite{Zhang2023, Sarkar2023, Kurumaji2025, Zhang2025, Gabriele2025, Sarkar2026}.   

We highlight key physical consequences of HNL antiferromagnets. In particular, in most HNL antiferromagnets, the constraints imposed by achiral symmetry and broken time-reversal symmetry make the third-order anomalous Hall effect the leading nonvanishing Hall response arising from Berry curvature multipoles.
More recently, nonlinear Hall effects arising from quantum metric dipoles have been intensively investigated theoretically \cite{Niu2014, Di2021, Yang2021_qmd, Das2023, Yan2024} and experimentally explored in several material platforms \cite{Suyang2023, Wang2023, Zhao2025}.
HNL antiferromagnets may also offer a promising topological platform for realizing sizable intrinsic nonlinear Hall responses enabled by their internal antiferromagnetic order rather than external magnetic fields.

Symmetry-controlled splitting of HNLs offers a direct pathway to generating anomalous Hall responses and topological chiral edge states.
Importantly, first-principles calculations have predicted strain-induced anomalous Hall and Nernst effects in the HNL antiferromagnet Mn$_3$GaN \cite{Samathrakis2020,Zhou2020}. 
In this case, strain lowers the symmetry from the inadmissible achiral MPG $\bar{3}m$ to the admissible MPG $\bar{3}$ and splits the HNLs. 
The coexistence of chiral edge modes and bulk conduction channels can serve as a characteristic signature of split HNLs.
One possible strategy to distinguish bulk and edge conduction channels is to incorporate decoherence effects into a six-terminal transport analysis \cite{sun2025}.
Notably, split HNLs can span a broad energy range around the Fermi energy, potentially providing a wide energy window for observing and manipulating these topological phenomena.
This suggests that engineering split HNLs broadens the routes to realizing anomalous Hall effects in antiferromagnetic crystals \cite{Chen2014, Kübler_2014, Nakatsuji2015, Ajaya2016, Yan2017, Gurung2019, Smejkal2022}. 

\section{Methods}  
\subsection*{\texorpdfstring{$\bm{k}\cdot\bm{p}$}{k.p} Hamiltonian construction} 
To determine the HNL directions emanating from high symmetry momenta, we construct the $\bm{k}\cdot\bm{p}$ Hamiltonians for two-dimensional IR coreps of inadmissible achiral MPGs. 
For each MPG, we impose the symmetry constraints on the two-band Hamiltonian and retain the lowest-order terms relevant to the protected degeneracies. 
The resulting Hamiltonians, together with the corresponding HNL types, are listed in Tables~\ref{tb_kp_colorless} and \ref{tb_kp_bnw}.

\begin{table*}[htbp]  		 
	\centering 
	\caption{\textbf{The $\bm{k}\cdot\bm{p}$ Hamiltonians for all 13 colorless inadmissible achiral magnetic point groups admitting Heesch nodal lines.} 
	The notation for two-dimensional double-valued irreducible corepresentations (2D IR coreps) follows the Bilbao Crystallographic Server Database \cite{Bilbao}. 
	The $\bm{k}\cdot\bm{p}$ Hamiltonians are expanded near degenerate high symmetry momenta with the corresponding little magnetic point groups (MPGs), keeping the lowest-order terms needed to display the Heesch nodal lines (HNLs). 
	The HNL types and the chiral subgroups hosting Heesch Weyl fermions (HWFs) are also listed.} \label{tb_kp_colorless}
	\begin{tabular}{l l l l l l}  
	\hline\hline
		MPG  & 2D IR coreps & $\bm{k}\cdot\bm{p}$ Hamiltonians & HNLs & Type & HWFs 
		\\ 
		\hline
		$mm2$ &$\bar{E}$ & $\alpha_{1}k_y \sigma_x+\alpha_{2}k_x\sigma_y$& $\hat{z}$ & I& ---
		\\ 
		$mmm$ &$\bar{E}_g,\bar{E}_u$ & $\begin{gathered}[t]
			\alpha_1k_yk_z\sigma_x+ \alpha_2k_xk_z\sigma_y
			+ \alpha_3k_xk_y\sigma_z
		\end{gathered}$ & $\hat{x}, \hat{y}, \hat{z}$  &I &$222$
		\\
		$4mm$ &$\bar{E}_2, \bar{E}_1$ 
		& $\alpha_1(k_y\sigma_x-k_x\sigma_y)$ & $\hat{z}$  &I & ---				
		\\  
		$\bar{4}2m$ &$\bar{E}_2, \bar{E}_1$ 
		& $\alpha_{1}(k_x\sigma_x-k_y\sigma_y)$ & $\hat{z}$ &I & $222$
		\\  
		$4/mmm$ & $\bar{E}_{2g(u)}, \bar{E}_{1g(u)}$  & $\alpha_{1}k_z(k_y\sigma_x-k_x\sigma_y) +\alpha_2k_xk_y(k_x^2-k_y^2)\sigma_z$ &  $\hat{x}, \hat{y}, \hat{z}, \pm\hat{x}\pm\hat{y}$ &I  &   $422, 222$
		\\
		$3m$ &$\bar{E}_1$ & $\alpha_1(k_y\sigma_x-k_x\sigma_y)$   &$\hat{z}$ &I  & ---
		\\
		$\bar{3}m$  &$\bar{E}_{1g}$, $\bar{E}_{1u}$ & $\begin{aligned}[t]
			&\alpha_{1}k_z(k_y\sigma_x-k_x\sigma_y) 
			\\&+ \alpha_2[(k_x^2+k_y^2)\sigma_x-2k_xk_y\sigma_y]
		\end{aligned}$ & $\hat{z}$&I  & $32$
		\\
		$6mm$  &$\bar{E}_3$   & $i\alpha_1 (k_+^3-k_-^3)\sigma_x +\alpha_2(k_+^3+k_-^3)\sigma_y$   &  $\hat{z}$ &I & ---
		\\  
		&$\bar{E}_1, \bar{E}_2$ &   $\alpha_1(k_y\sigma_x-k_x\sigma_y)$ &  $\hat{z}$ &I  & ---
		\\  
		$\bar{6}m2$ &$\bar{E}_3$ &  $\begin{aligned}[t]
			&[\alpha_1+\alpha_2(k_x^2+k_y^2)+\alpha_3k_z^2]k_z\sigma_y 
			\\&+ i\alpha_4k_y(3k_x^2-k_y^2)\sigma_z 
		\end{aligned}$ &$\hat{x},C_3\hat{x},C_3^2\hat{x}$ &I & $32$
		\\  
		&$\bar{E}_1,\bar{E}_2$ & $\begin{aligned}[t]
			&\alpha_1k_z(k_y\sigma_x-k_x\sigma_y)+ \alpha_2k_y(3k_x^2-k_y^2)\sigma_z \\&+\alpha_3k_z[2k_xk_y\sigma_x +(k_x^2-k_y^2)\sigma_y]	\end{aligned}$
		&  $\hat{x},C_3\hat{x},C_3^2\hat{x},\hat{z}$  &I &---
		\\  
		$6/mmm$ &$\bar{E}_{3g}, \bar{E}_{3u}$ & $\begin{aligned}[t]
			&k_z[\alpha_1k_y(3k_x^2-k_y^2)\sigma_x +\alpha_2k_x(k_x^2-3k_y^2)\sigma_y]
			\\&+\alpha_3k_xk_y (3k_x^2-k_y^2) (k_x^2-3k_y^2)\sigma_z
		\end{aligned}$ & $\begin{aligned}[t]&\hat{x},C_6\hat{x},C_3\hat{x},\\&\hat{y},C_3^2\hat{x},C_6^5\hat{x},\hat{z}\end{aligned}$&I & $622$
		\\  
		&$\bar{E}_{1g(u)}, \bar{E}_{2g(u)}$ & $\begin{aligned}[t]
			&[\alpha_1+\alpha_2k_z^2+\alpha_3(k_x^2+k_y^2)]k_z(k_y\sigma_x-k_x\sigma_y)
			\\&+\alpha_4k_xk_y (3k_x^2-k_y^2) (k_x^2-3k_y^2)\sigma_z
		\end{aligned}$ &  $\begin{gathered}[t] 
			\hat{x},C_6\hat{x},C_3\hat{x},\hat{y},C_3^2\hat{x},C_6^5\hat{x},\hat{z}
		\end{gathered}$ &I &  ---
		\\ 
		$m\bar{3}$ &$\begin{gathered}[t]
			\bar{E}_{g(u)},  {}^1\bar{F}_{g(u)}, {}^2\bar{F}_{g(u)}
		\end{gathered}$ & $ \alpha_1 (k_yk_z\sigma_x+k_xk_z\sigma_y+k_xk_y\sigma_z)$   & $\hat{x},\hat{y},\hat{z}$ &I & $23, 222$ 
		\\ 
		$\bar{4}3m$ &$\bar{E}_{1},\bar{E}_{2}$ & $\begin{aligned}[t]
			\alpha_1 [&k_x(k_y^2-k_z^2)\sigma_x +k_y(k_z^2-k_x^2)\sigma_y\\&+k_z(k_x^2-k_y^2)\sigma_z]
		\end{aligned}$   & $\begin{aligned}[t]
			\hat{x},\hat{y},\hat{z}, \pm\hat{x}\pm\hat{y}\pm\hat{z}
		\end{aligned}$ &I & $23, 222$
		\\ 
		$m\bar{3}m$ &$\bar{E}_{1g(u)}, \bar{E}_{2g(u)}$ &$\begin{aligned}[t]
			&\alpha_1[k_yk_z(k_y^2-k_z^2)\sigma_x\\&+k_zk_x(k_z^2-k_x^2)\sigma_y+k_xk_y(k_x^2-k_y^2)\sigma_z]
		\end{aligned}$ & $\begin{aligned}[t]
			&\hat{x}, \hat{y}, \hat{z},\pm\hat{x}\pm\hat{y}\pm\hat{z} ,
			\\& \pm\hat{x}\pm\hat{y}, 
			\pm\hat{y}\pm\hat{z}, \pm\hat{z}\pm\hat{x} 
		\end{aligned}$ &I & $\begin{aligned}[t] &432,23,32,\\ & 422, 222 \end{aligned}$
		\\ 
		\hline\hline 
	\end{tabular}
\end{table*}

\begin{table*}[htbp]  
	\renewcommand{\arraystretch}{1}
	\centering
	\caption{\textbf{The $\bm{k}\cdot\bm{p}$ Hamiltonians for all 14 black-and-white inadmissible achiral magnetic point groups admitting Heesch nodal lines.} 
	The notation for two-dimensional double-valued irreducible corepresentations (2D IR coreps) follows the Bilbao Crystallographic Server Database \cite{Bilbao}. 
	The $\bm{k}\cdot\bm{p}$ Hamiltonians are expanded near degenerate high symmetry momenta with the corresponding little magnetic point groups (MPGs), retaining the lowest-order terms needed to capture the Heesch nodal lines (HNLs). 
	The HNL types and the chiral subgroups hosting Heesch Weyl fermions (HWFs) are also listed.}
	\label{tb_kp_bnw}	
	\begin{tabular}{l l l l l l} 
		\hline\hline
		MPG  &2D IR coreps & $\bm{k}\cdot\bm{p}$ Hamiltonians & HNLs &  Type & HWFs	
		\\ \hline
		$\bar{4}^\prime$ &$\leftindex^{2}{\bar{E}}\leftindex^{1}{\bar{E}}$ & $\begin{gathered}
			\alpha_1(k_y\sigma_x +k_x\sigma_y)
		\end{gathered}$ & $\hat{z}$  &II & ---
		\\ 
		$4^\prime/m$ &$\leftindex^{2}{\bar{E}}_g\leftindex^{1}{\bar{E}}_g, \leftindex^{2}{\bar{E}}_u\leftindex^{1}{\bar{E}}_u$  & $\begin{aligned}[t]
			&k_z(\alpha_1k_y+\alpha_2k_x)\sigma_x  +k_z(\alpha_1k_x-\alpha_2k_y)\sigma_y \\&+[\alpha_3k_xk_y+\alpha_4(k_x^2-k_y^2)]\sigma_z\end{aligned}$ & $\hat{z}$ &II  	& $4^\prime$
		\\
		$4^\prime m^\prime m $  &$\bar{E}$ 
		& $\alpha_1(k_y\sigma_x -k_x\sigma_y)$ & $\hat{z}$ &I & $4^\prime$
		\\  
		$\bar{4}^\prime 2m^\prime$ &$\bar{E}$ 
		& $\alpha_1(k_x\sigma_x -k_y\sigma_y)$ & $\hat{z}$ & II &$222$
		\\  
		$\bar{4}^\prime 2^\prime m$ &$\bar{E}$ 
		& $\alpha_1(k_x\sigma_x -k_y\sigma_y)$ & $\hat{z}$  & I(II)  &---
		\\  
		$4^\prime /mm^\prime m$ & $\bar{E}_{g}, \bar{E}_{u}$  & $\begin{gathered}
			\alpha_1k_z(k_y\sigma_x -k_x\sigma_y) +\alpha_2(k_x^2-k_y^2)\sigma_z
		\end{gathered}$ &  $\hat{z}, \pm\hat{x}\pm\hat{y}$  &I(II)  &  $4^\prime$
		\\
		$\bar{6}^\prime$ &$\leftindex^{1}{\bar{E}}\leftindex^{2}{\bar{E}}$ & $\alpha_1 [(k_x^2-k_y^2)\sigma_x-2k_xk_y\sigma_y]  +\alpha_2[2k_xk_y\sigma_x+(k_x^2-k_y^2)\sigma_y]$  &$\hat{z}$ & II &--- 
		\\
		$6^\prime/m^\prime$ &$\leftindex^{1}{\bar{E}}_{g(u)}\leftindex^{2}{\bar{E}}_{g(u)}$ & 
		$\alpha_1[(k_x^2-k_y^2)\sigma_x-2k_xk_y\sigma_y]+\alpha_2[2k_xk_y\sigma_x+(k_x^2-k_y^2)\sigma_y]$ & $\hat{z}$& II &   $6^\prime$
		\\ 
		$6^\prime mm^\prime$ &$\bar{E}_1$   & $\alpha_1(k_y\sigma_x-k_x\sigma_y)$   & $\hat{z}$ &I &$6^\prime$
		\\  
		$\bar{6}^\prime m^\prime2$ &$\bar{E}_1$ &  $\alpha_1[(k_x^2-k_y^2)\sigma_x-2k_xk_y\sigma_y]$ &$\hat{z}$ &II &$32$
		\\  
		$\bar{6}^\prime m2^\prime$ &$\bar{E}_1$ &  $\alpha_1[2k_xk_y\sigma_x+(k_x^2-k_y^2)\sigma_y]$ &$\hat{z}$ & I(II) &---
		\\  
		$6^\prime/m^\prime mm^\prime $ &$\bar{E}_{1g}, \bar{E}_{1u}$ & $\alpha_1[2k_xk_y\sigma_x+(k_x^2-k_y^2)\sigma_y]$ & $\hat{z}$& I(II) & $6^\prime22^\prime$
		\\ 
		$\bar{4}^\prime3m^\prime$  &$\bar{E},\leftindex^{1}{\bar{F}}, \leftindex^{2}{\bar{F}}$ & $\alpha_1 (k_yk_z\sigma_x+k_zk_x\sigma_y+k_xk_y\sigma_z) $  & $\hat{x},\hat{y},\hat{z}$ & II & $23, 222$
		\\  
		$m\bar{3}m^{\prime}$ 
		&$\bar{E}_{g(u)},\leftindex^{1}{\bar{F}}_{g(u)}, \leftindex^{2}{\bar{F}}_{g(u)}$ &$\alpha_1 (k_yk_z\sigma_x+k_zk_x\sigma_y+k_xk_y\sigma_z)$ & $\hat{x},\hat{y},\hat{z}$ & I(II) &$
		4^\prime32^\prime, 23,  4^\prime22^\prime, 222$  
		\\ 
		\hline\hline
	\end{tabular}
\end{table*}

\subsection{Tight-binding Hamiltonian of Mn$_3$GaN}
In this section, we construct a tight-binding Hamiltonian for Mn$_3$GaN in the basis of Mn s orbitals. There are three Mn atoms per unit cell, located at $\bm{r}_{{\rm Mn},1}=-(1,0,0)a/2$, $\bm{r}_{{\rm Mn},2}=-(0,1,0)a/2$ and $\bm{r}_{{\rm Mn},3}=-(0,0,1)a/2$, where $a$ is the lattice constant. The total Hamiltonian includes three parts:
\begin{equation}\label{eq_H}
	H = H_0 + H_J +H_{\rm SOC}.
\end{equation}

The first part $H_0$ describes nearest-neighbor hopping between adjacent Mn $s$ orbitals,  
\begin{equation}\label{eq_H0}
\begin{aligned}
H_0 =&-\epsilon_0\sum_{\bm{k},\alpha}
\hat{c}_{\bm{k},\alpha}^{\dagger}\hat{c}_{\bm{k},\alpha}\\
&+\sum_{\bm{k},\alpha\neq\beta}
\left(2t_1\cos\phi_{\alpha\beta,1}
+2t_2\cos\phi_{\alpha\beta,2}\right)
\hat{c}_{\bm{k},\alpha}^{\dagger}\hat{c}_{\bm{k},\beta}.
\end{aligned}
\end{equation}
Here, $\alpha,\beta\in\{1,2,3\}$ label the kagome sublattices. $t_1$ and $t_2$ are real parameters denoting intra- and inter-kagome-layer nearest-neighbor hoppings, respectively.  $\phi_{\alpha\beta,1} =  \bm{k} \cdot(\bm{r}_{{\rm Mn},\alpha}-\bm{r}_{{\rm Mn},\beta})$ and  $\phi_{\alpha\beta,2} =  \bm{k} \cdot(\bm{r}_{{\rm Mn},\alpha}+\bm{r}_{{\rm Mn},\beta})$. 

The second term, $H_J$, represents the onsite exchange coupling between the itinerant $s$ electrons and the localized Mn $d$ moments,
\begin{equation}\label{eq_HJ}
	H_J= J \sum_{\bm{k}, \alpha, \gamma\tau} \bm{m}_{\alpha}\cdot\bm{\sigma}_{\gamma\tau} \hat{c}_{\bm{k},\alpha,\gamma}^{\dagger}\hat{c}_{\bm{k},\alpha,\tau}  .
\end{equation}
Here, $\bm{\sigma}$ is the vector of Pauli matrices and  $\bm{m}_{\alpha}$ are the local magnetic moments of the three Mn atoms in each unit cell, defined as $\bm{m}_1 =(0,1,-1)/\sqrt{2}$, $\bm{m}_2 =(-1,0,1)/\sqrt{2}$ and $\bm{m}_3 = (1,-1,0)/\sqrt{2}$.

The spin-dependent hopping Hamiltonian associated with SOC is
\begin{equation}\label{eq_HSOC}
	H_{\rm SOC} = \sum_{\bm{k}, \alpha\neq\beta,\gamma\tau} 2\bm{t}_{\beta,\alpha}^{\rm SOC}\cdot \bm{\sigma}_{\gamma\tau} \cos\phi_{\alpha\beta,1} \hat{c}_{\bm{k},\alpha,\gamma}^{\dagger}\hat{c}_{\bm{k},\beta,\tau},
\end{equation}
with 
\begin{equation}
	\begin{aligned}
		\bm{t}_{2,1}^{\rm SOC} &= \left(t_{\rm SOC,1} e^{-i\phi_{\rm SOC}}, -t_{\rm SOC,1} e^{i\phi_{\rm SOC}}, -it_{\rm SOC,2}\right),\\
		\bm{t}_{1,3}^{\rm SOC} &= \left( -t_{\rm SOC,1} e^{i\phi_{\rm SOC}}, -it_{\rm SOC,2}, t_{\rm SOC,1} e^{-i\phi_{\rm SOC}}\right),\\
		\bm{t}_{3,2}^{\rm SOC} &= \left(-it_{\rm SOC,2}, t_{\rm SOC,1} e^{-i\phi_{\rm SOC}},  -t_{\rm SOC,1} e^{i\phi_{\rm SOC}}\right),	
	\end{aligned}
\end{equation}
and $\bm{t}_{\alpha,\beta}^{\rm SOC} = \left(\bm{t}_{\beta,\alpha}^{\rm SOC}\right)^*$. $t_{\rm SOC,1}$, $t_{\rm SOC,2}$ and $\phi_{\rm SOC}$ are all real. 
A nontrivial phase $e^{i\phi_{\rm SOC}}\neq 1$ and a finite imaginary hopping $-it_{\rm SOC,2}$ can generate nontrivial phases along nearest-neighbor Mn bonds within the same kagome layer, thereby producing effective fluxes through closed hopping loops and locally nonzero Berry curvature. 

To reduce the MPG from $\bar{3}m$ to $32$, one can add the following perturbation to break achiral symmetries
\begin{equation} \label{eq_dH32}
	\delta H_{0}^{32}  = \sum_{\bm{k},\alpha\beta} 2\nu_{\alpha\beta}\left(\delta t_1 \sin\phi_{\alpha\beta,1} + i \delta t_2 \sin\phi_{\alpha\beta,2}  \right) \hat{c}_{\bm{k},\alpha}^\dagger \hat{c}_{\bm{k},\beta},
\end{equation}
where $\nu_{\alpha\beta}$ is antisymmetric with respect to its sublattice indices and $\nu_{12}=\nu_{23}=\nu_{31}=1$. $\delta t_1$ and $\delta t_2$ are real. 

The $t_{\rm SOC,1}$ and $\delta t_1$ terms break time-reversal symmetry and encode the interplay between magnetic order and nearest-neighbor hopping.

\subsection{Higher-order anomalous Hall effect}
The intrinsic anomalous Hall conductivity can be evaluated by integrating the Berry curvature over the Brillouin zone
\begin{equation}
	\sigma_{\alpha\beta} = -\frac{e^2}{\hbar} \int \frac{d^3\bm{k}}{(2\pi)^3} \sum_n f[\epsilon_n(\bm{k})-\mu] \Omega_{n,\alpha\beta} (\bm{k}),
\end{equation}
where $f[\epsilon_n(\bm{k})-\mu]$ is the Fermi–Dirac distribution, $\mu$ is the chemical potential, $n$ is the band index, and $\epsilon_{n}(\bm{k})$ is the energy dispersion. The band-resolved Berry curvature is given by 
\begin{equation}
	\Omega_{n,\alpha\beta} (\bm{k}) 
	= -2 {\rm Im} \sum_{m\neq n} \frac{\bra{n \bm{k}} v_\alpha\ket{m \bm{k}} \bra{m \bm{k}} v_\beta\ket{n \bm{k}} }{[\epsilon_m(\bm{k}) -\epsilon_n(\bm{k})]^2},
\end{equation}
where $(\alpha,\beta,\gamma)$ is cyclic,
$\Omega_{n,\alpha\beta}=\epsilon_{\alpha\beta\gamma}\Omega_{n,\gamma}$, and $v_\alpha=\partial_{k_\alpha}H$.

The Berry curvature dipole is defined as
\begin{equation}	\label{eq_dipole}
	D_{\alpha\beta} = \int\frac{d^3\bm{k}}{(2\pi)^3}  \sum_n f[\epsilon_n(\bm{k})-\mu]\partial_{k_\alpha}\Omega_{n,\beta}(\bm{k}),
\end{equation}
where $\alpha,\beta=\{x,y,z\}$. Under an AC electric field, a second-order Hall signal proportional to the relaxation time $\tau$ can be generated due to the finite Berry curvature dipole moment \cite{PhysRevLett.115.216806}. Similarly, a third-order nonlinear Hall conductivity proportional to $\tau^2$ can be induced by a Berry curvature quadrupole \cite{PhysRevB.107.115142}
\begin{equation} 	\label{eq_quadrupole}
	Q_{\alpha\beta\gamma}  =  \int\frac{d^3\bm{k}}{(2\pi)^3} \sum_n f[\epsilon_n(\bm{k})-\mu]  \partial_{k_\alpha}\partial_{k_\beta}\Omega_{n,\gamma}(\bm{k}). 
\end{equation}

The Berry curvature behaves as a pseudovector. A symmetry operation $R$ can impose constraints on the Berry curvature, dipole and quadrupole:
\begin{equation}
	\bm{\Omega}(\bm{k}) = \pm \det{(R)} R^T\bm{\Omega}(\pm R\bm{k}),
\end{equation}  
\begin{equation}
	D_{\alpha\beta}
	= \det(R)\sum_{\alpha^{\prime},\beta^\prime}    D_{\alpha^{\prime}\beta^{\prime}} R_{\alpha^{\prime} \alpha}R_{\beta^\prime\beta},
\end{equation}
and 
\begin{equation}
	Q_{\alpha\beta\gamma}  
	= \pm \det(R)\sum_{\alpha^{\prime}\beta^{\prime}\gamma^\prime}    	Q_{\alpha^{\prime}\beta^{\prime}\gamma^{\prime}} R_{\alpha^{\prime} \alpha} R_{\beta^{\prime} \beta}R_{\gamma^\prime\gamma},
\end{equation}
respectively. Here, $+$($-$) is taken for unitary (antiunitary) operations. Based on the symmetry constraints, we give a full list of the Berry curvature dipole and quadrupole for all 27 inadmissible achiral MPGs that host HNLs in Supplementary Note 4.

When the principal axis is set to the [111] direction, the MPG $\bar{3}m$, generated by $C_{3,111}$,$C_{2,1\bar{1}0}$ and $I$, admits the following nonzero quadrupole components
\begin{equation}
	Q_{xyy} = Q_{yzz} = Q_{zxx} = -Q_{yxx} = -Q_{zyy}= -Q_{xzz},
\end{equation}
and
\begin{equation}
	Q_{xxy} = Q_{yyz} = Q_{zzx} = -Q_{yyx} = -Q_{zzy}= -Q_{xxz}.
\end{equation}
Its subgroup $32$ hosting HWFs admits the second-order Hall response induced by Berry curvature dipoles
\begin{equation}
	D_{xx} = D_{yy} = D_{zz},
\end{equation}
and
\begin{equation}
	D_{xy} = D_{yz} = D_{zx} = D_{yx} = D_{zy} = D_{xz}.
\end{equation}

\section{Data Availability} 
The data that support the findings of this study are available from the corresponding author upon reasonable request.

\section{Code Availability }
The codes generated during this study are available from the corresponding authors upon reasonable request.

\section{Acknowledgements}
\begin{acknowledgments}
We thank Wen-Bo Dai for helpful discussions.
K. T. L. acknowledges the support of the Ministry of Science and Technology, China, The New Cornerstone Foundation, and the Hong Kong Research Grants Council through Grants No. MOST23SC01-A, No. RFS2021-6S03, No. C6053-23G, No. AoE/P-701/20, AoE/P-604/25R, No. 16309223, No. 16311424 and No. 16300325.
\end{acknowledgments}
  
\section{Author Contributions }
K.T.L. conceived the project.
X.-Y.G. carried out the investigation and wrote the manuscript under the guidance of Z.-T.S. and K.T.L.
C.-Y.C. contributed to the identification of candidate materials and performed the first-principles calculations.
All authors contributed to the scientific discussions and manuscript revisions.

\section{Competing interests}
The authors declare no competing interests.

\clearpage
\onecolumngrid 
\begin{center}
	\textbf{\large Supplementary Information -- Heesch Nodal Lines in Inadmissible Achiral Antiferromagnets}\\[.2cm]
	Xing-Yao Guo, Chung-Yuen Chan, Zi-Ting Sun,$^{*}$ and Kam Tuen Law$^{\dagger}$\\[.1cm]
	{\itshape  Department of Physics, Hong Kong University of Science and Technology, Clear Water Bay, Hong Kong, China} 
\end{center} 

\setcounter{equation}{0}
\setcounter{section}{0}
\setcounter{figure}{0}
\setcounter{table}{0}
\setcounter{page}{1}

\renewcommand{\theequation}{S\arabic{equation}}
\renewcommand{\theHequation}{SM.\arabic{equation}}
\renewcommand{\thefigure}{\arabic{figure}}
\renewcommand{\theHfigure}{SM.\arabic{figure}}
\renewcommand{\thetable}{\arabic{table}}
\renewcommand{\theHtable}{SM.\arabic{table}}

\setcounter{secnumdepth}{1}
\renewcommand{\thesection}{Supplementary Note \arabic{section}}
\renewcommand{\theHsection}{SM.\arabic{section}}

\makeatletter
\def\fnum@figure{\textbf{Supplementary Figure~\thefigure}}
\def\fnum@table{\textbf{Supplementary Table~\thetable}}
\makeatother
\section{First-principles calculations} 
Throughout this work, the \textit{Vienna Ab initio Simulation Package} (VASP)~\cite{SM_KRESSE199615} with the projector-augmented wave method~\cite{SM_PhysRevB.50.17953} and the Perdew--Burke--Ernzerhof (PBE) exchange-correlation functional in the generalized-gradient approximation~\cite{SM_PhysRevB.28.1809,SM_PhysRevLett.77.3865} were used to perform the first-principles calculations~\cite{SM_PhysRev.136.B864}. Information about calculated materials such as the lattice structures was obtained mainly from the MAGNDATA database of the Bilbao server~\cite{SM_MAGNDATA_I,SM_MAGNDATA_II}. Onsite Coulomb interactions were included with the DFT+U method, where we set $U_{\rm eff} =U-J =4~$eV for Mn $d$-orbital electrons in Mn$_3$GaN, $U_{\rm eff}=3~$eV for Mn $d$-orbital electrons in MnTe, and $U_{\rm eff}=0.8~$eV for Cr $d$-orbital electrons in CrSb.

\section{Two-dimensional irreducible corepresentations in inadmissible magnetic point groups}  
In this note, we explain how two-dimensional double-valued irreducible corepresentations (2D IR coreps) arise in inadmissible magnetic point groups (MPGs) lacking both time-reversal and parity-time symmetries. 
These two-dimensional IR coreps enforce twofold degeneracies at high symmetry momenta respecting the corresponding magnetic little groups, which serve as the starting points from which Heesch nodal lines (HNLs) emanate. 

Under a symmetry operation $\hat{g}\in G_{\bm{k}}$, the wavefunction basis $\ket{\psi}$ transforms according to the corepresentation matrix $D(\hat{g})$,
\begin{equation}
	P_{\hat{g}} \ket{\psi} = \ket{\psi}D(\hat{g}) .
\end{equation}
In general, $\ket{\psi}$ can be factorized into a momentum-dependent part and an internal part $\ket{m}$ labeled by the magnetic quantum number along the principal axis, where $m$ is half-integer in electronic systems. 
At the high symmetry point $\bm{k}_0$, the momentum part remains unchanged under operations in the little group $G_{\bm{k}_0}$, since these operations leave the momentum $\bm{k}_0$ invariant up to a reciprocal lattice vector,
$G_{\bm{k}_0}=\left\{\hat{g}\in G \mid \hat{g}\bm{k}_0 \equiv \bm{k}_0+\bm{G}\right\}$.
Therefore, for simplicity, one can focus on
\begin{equation}
	P_{\hat{g}} \ket{m} = \ket{m}D(\hat{g}).
\end{equation}

For a unitary point group symmetry operation, the corepresentation matrix $D(\hat{g})$ is unitary and can be diagonalized by a basis transformation $\ket{m'}=\ket{m}S$, such that $P_{\hat{g}}\ket{m'}=\ket{m'}\Lambda$, with
\begin{equation}\label{SM_eq_diag_u}
	S^{-1}D(\hat{g})S = \Lambda .
\end{equation}
For colorless groups, a two-dimensional IR corep exists only when the representation matrices of at least two symmetry operations, typically the group generators, cannot be simultaneously diagonalized by the same unitary transformation matrix $S$ via Eq.~\eqref{SM_eq_diag_u}. 
This condition is satisfied by all inadmissible colorless groups. 
In Supplementary Table~\ref{tb_BCDBCQ}, the generators of colorless inadmissible achiral groups always involve two rotations or rotoinversions along different axes, accounting for the presence of two-dimensional IR coreps. 

For black-and-white MPGs containing antiunitary operations, e.g., $\hat{T}\hat{g}$, the corepresentation matrix $D(\hat{T}\hat{g})$ can be written as the product of a matrix $M$ and the complex conjugation operator $K$, $D(\hat{T}\hat{g})=MK$. 
The two-dimensional corepresentation is reducible in the new basis $\ket{m'}=\ket{m}S$ if a matrix $S$ can be found such that the matrix $M$ can be diagonalized via
\begin{equation}\label{SM_eq_diag_au}
	S^{-1}M S^*= \Lambda.
\end{equation}  
Therefore, a single antiunitary generator can, in certain cases, support a two-dimensional IR corep. 
This is exemplified by the black-and-white group $\bar{4}'$, generated by ${T}IC_{4z}$. 
In the basis $\ket{\pm 1/2}$, its corepresentation matrix is 
$D({T}IC_{4z}) = \begin{pmatrix}0&i\\1&0\end{pmatrix}K$, 
which cannot be diagonalized via Eq.~\eqref{SM_eq_diag_au}. 
The defining characteristic of an inadmissible black-and-white group is that no transformation $S$ can simultaneously diagonalize the corepresentations of all group elements, including both unitary and antiunitary operations, by Eqs.~\eqref{SM_eq_diag_u} and \eqref{SM_eq_diag_au}, respectively. 
This irreducibility mandates the existence of a two-dimensional IR corep.
 
Moreover, in contrast to grey groups where Kramers degeneracy universally enforces twofold degeneracy, inadmissible colorless or black-and-white MPGs can host both 1D and 2D IR coreps in some cases, depending on the angular momentum of the basis states. 
For instance, the inadmissible group $32$ is generated by $C_{3z}$ and $C_{2x}$. 
In the $\ket{\pm 1/2}$ basis, the matrices $D(C_{3z})$ and $D(C_{2x})$ cannot be diagonalized simultaneously, resulting in the 2D IR corepresentation $\bar{E}_1$. 
However, for states with higher angular momentum $\ket{\pm 3/2}$, $D(C_{3z})$ becomes proportional to the identity matrix. 
This allows for simultaneous diagonalization, reducing the two-dimensional corep into two distinct one-dimensional IR coreps (${}^2\bar{E}$ and ${}^1\bar{E}$). 
Similar arguments explain the coexistence of 2D IR and 1D IR coreps in related achiral groups such as $3m$ and $\bar{3}m$, where the additional achiral symmetry operations introduce only phase factors and do not affect the reducibility.

\section{Symmetry-enforced Heesch nodal lines}
In this note, we use a two-band model to demonstrate how nodal lines emanate from a degenerate high symmetry momentum protected by inadmissibility for $m=\pm 1/2$ fermions.
Achiral symmetries impose constraints on spin polarization, which generically lead to the formation of symmetry-protected HNLs in these systems.

The Hamiltonian near  $\bm{k}_0$ can be generically expressed as
\begin{equation}
	H(\bm{k}) = f_0(\bm{k})\sigma_0+ \bm{f}(\bm{k})\cdot  \bm{\sigma},
\end{equation}
Here, $\bm{k}$ denotes the momentum measured relative to the high symmetry momentum $\bm{k}_0$. $\bm{k}= \left[k_+,k_-,k_z\right]^T$ and  $\bm{\sigma}= \left[\sigma_+,\sigma_-,\sigma_z\right]^T$ are all written in the helical basis for the following analysis of $n$-fold rotation operations, with $k_\pm = k_x \pm ik_y$ and $\sigma_\pm = \sigma_x \pm i\sigma_y$. The coefficients are $\bm{f}(\bm{k}) = \left[f_+(\bm{k}),f_-(\bm{k}),f_z(\bm{k})\right]^T$. The term $\bm{f}(\bm{k}) \cdot \bm{\sigma}$ is the dot product and denotes the spin dependent term due to spin-orbit coupling or magnetic order.

The symmetry operation $\hat{g}\in G_{\bm{k}_0}$ imposes the following constraint on the Hamiltonian near $\bm{k}_0$,
\begin{equation}
	H(\bm{k}) = D_{1/2}^{-1}(\hat{g}) H(\hat{g}\bm{k})D_{1/2}(\hat{g}).
\end{equation}
For the spin-dependent part, this gives
\begin{equation}
	\begin{aligned}
		\bm{f}(\bm{k}) \cdot  \bm{\sigma} 
		= \bm{f}(\hat{g}\bm{k}) \cdot   D_{1/2}^{-1}(\hat{g}) \bm{\sigma} D_{1/2}(\hat{g}) 
		=\bm{f}(\hat{g}\bm{k}) \cdot \left[  \det(\hat{g})  \hat{g}\bm{\sigma}\right] 
		= \left[\det(\hat{g})  \hat{g}^T \bm{f}(\hat{g}\bm{k})\right] \cdot \bm{\sigma}  .
	\end{aligned}
\end{equation}
Therefore, 
\begin{equation}\label{SM_eq_fk}
	\bm{f}(\bm{k})  = \det(\hat{g})  \hat{g}^T \bm{f}(\hat{g}\bm{k}) .
\end{equation}
In the following, we set the $\hat{z}$ axis as the principal rotation axis without loss of generality.

\begin{figure} [h]
		\includegraphics[width=0.9\columnwidth]{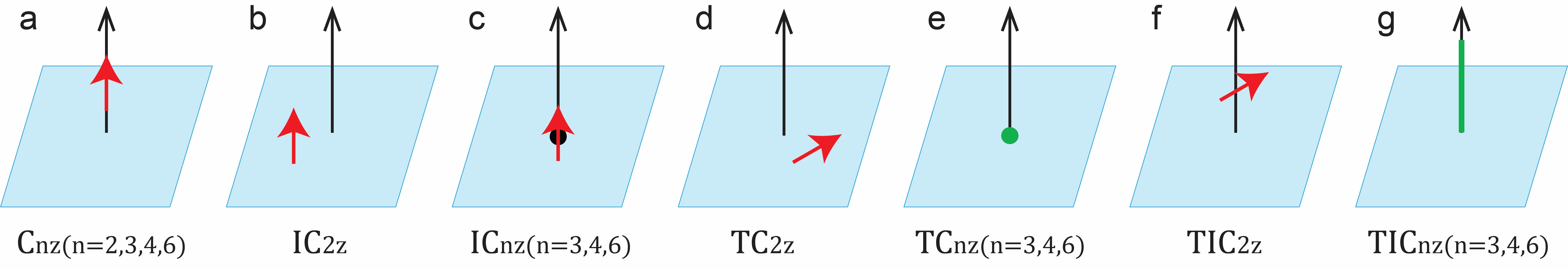}
		\caption{\textbf{Spin-polarization constraints for different symmetry operations.} 
		The black arrow represents the rotation axis, and the red arrows denote the allowed spin-polarization directions. 
		The black point denotes the origin, while the green dots and lines represent symmetry-enforced degenerate points and lines, respectively.}
		\label{fig_spin_constrain}
\end{figure}
\textbf{Constraints on spin polarization from unitary operations.}
The symmetry transformation matrix for an $n$-fold rotation $C_n$ or a rotoinversion $IC_n$ operation is 
\begin{equation}
	\hat{g}=  \eta \begin{pmatrix} e^{i2\pi/n} & & \\ & e^{-i2\pi/n} &\\ & & 1	\end{pmatrix}, 
\end{equation}
where $\eta=1$ for pure rotations and $\eta=-1$ for rotoinversions. Applying the general rule in Eq.~\eqref{SM_eq_fk} imposes
\begin{equation}
	f_\pm(k_+,k_-,k_z) = e^{\pm i2\pi/n}f_\pm(\eta e^{i2\pi/n}k_+, \eta e^{-i2\pi/n}k_-, \eta k_z),
\end{equation}
\begin{equation}
	f_z(k_+,k_-,k_z) = f_z(\eta e^{i2\pi/n}k_+, \eta e^{-i2\pi/n}k_-, \eta k_z).
\end{equation} 
As a result, we have the following specific constraints on spin polarization.\\
(a) For $n$-fold rotations $C_n$ with $n=2,3,4,6$, on the rotation axis ($k_+=k_-=0$), Eq.~\eqref{SM_eq_fk} implies $f_\pm = 0$, while $f_z$ remains generically nonzero. This indicates that only spin polarization parallel (or antiparallel) to the rotation axis is permitted for momenta on the axis (Supplementary Fig.~\ref{fig_spin_constrain}a). \\
(b) For the mirror symmetry $M_z = IC_{2z}$ ($n=2$), in the mirror plane ($k_z=0$), $f_\pm = 0$, while generally $f_z \neq 0$. Thus, only spin polarization perpendicular to the mirror plane is allowed for momenta within the plane (Supplementary Fig.~\ref{fig_spin_constrain}b).\\
(c) For $IC_n$ with $n=3,4,6$, at the origin ($\bm{k}=0$), the constraints force $f_\pm = 0$ but allow $f_z \neq 0$, meaning only out-of-plane spin polarization is permitted at the origin (Supplementary Fig.~\ref{fig_spin_constrain}c).

\textbf{Constraints on spin polarization from antiunitary operations.}
We now consider the constraints imposed by antiunitary symmetries combining time reversal with an $n$-fold operation ($TC_n$ or $TIC_n$). 
The transformation matrix in the helical basis for these operations is
\begin{equation}
	\hat{g}=  \eta \begin{pmatrix} &-e^{-i2\pi/n}  & \\ -e^{i2\pi/n}&  &\\ & & -1	\end{pmatrix}, 
\end{equation}
where $\eta=1$ for $TC_n$ and $\eta=-1$ for $TIC_n$. 
The antiunitary operation exchanges the coefficients $f_+$ and $f_-$. Applying the corresponding antiunitary symmetry constraint twice to $f_{\pm}$ and once to $f_z$, we obtain
\begin{equation} \label{SM_eq_au_f}
	f_\pm(k_+,k_-,k_z) = e^{\mp i4\pi/n}f_\pm(e^{i4\pi/n}k_+, e^{-i4\pi/n}k_-, k_z).
\end{equation}
\begin{equation} \label{SM_eq_au_fz}
	f_z(k_+,k_-,k_z) = -f_z(-\eta e^{-i2\pi/n}k_-, -\eta e^{i2\pi/n}k_+, -\eta k_z).
\end{equation}
Therefore, we have the following specific constraints on spin polarization.\\
(d) For $TC_{2z}$, on the $k_z=0$ plane, Eqs.~\eqref{SM_eq_au_f} and \eqref{SM_eq_au_fz} force $f_z =0$, while $f_\pm$ can be nonzero, implying that only in-plane spin polarization is allowed (Supplementary Fig.~\ref{fig_spin_constrain}d).\\
(e) For $TC_{nz}$ ($n =3,4,6$), at the origin ($\bm{k}=0$), $f_z=0$ and $f_\pm = 0$, indicating a degenerate point (Supplementary Fig.~\ref{fig_spin_constrain}e).\\
(f) For $TIC_{2z}$, on the rotation axis ($k_+=k_-=0$), $f_z=0$ but $f_\pm \neq 0$, again allowing only in-plane spin polarization (Supplementary Fig.~\ref{fig_spin_constrain}f).\\
(g) For $TIC_{nz}$ ($n =3,4,6$), on the rotation axis ($k_+=k_-=0$), the constraints force $f_z=0$ and $f_\pm = 0$. This implies a line of degeneracy along the $\hat{z}$ direction (Supplementary Fig.~\ref{fig_spin_constrain}g).

The interplay of these constraints provides a mechanism for generating HNLs. 
For example, when two mirror planes intersect, each mirror symmetry allows only spin polarization perpendicular to its own mirror plane. 
On their intersection line, these two constraints become incompatible and force all spin-dependent terms to vanish, thereby enforcing a type I HNL. 
By contrast, admissible achiral colorless MPGs have no mirror symmetry or only a single mirror symmetry, explaining why nodal lines are not enforced in these groups. 
In black-and-white MPGs, Supplementary Fig.~\ref{fig_spin_constrain}g similarly implies a degenerate line along the $\hat{z}$ direction, constituting a type II HNL.

\section{Berry curvature dipoles and quadrupoles in inadmissible achiral magnetic point groups}
Based on the symmetry constraints shown in the Methods section of the main text, we present the independent nonzero Berry curvature dipole and quadrupole components for all 27 inadmissible achiral MPGs hosting HNLs in Supplementary Table \ref{tb_BCDBCQ}. Here, \(Q_{\alpha\beta\gamma}=Q_{\beta\alpha\gamma}\).
\begin{table*}[htbp] 
	\centering
	\renewcommand{\arraystretch}{1.1}
	\caption{The independent nonzero Berry curvature dipole $D_{\alpha\beta}$ and quadrupole $Q_{\alpha\beta\gamma}$ for all 27 inadmissible achiral magnetic point groups (MPGs) hosting Heesch nodal lines. The generators of the MPGs are listed to help identify the $x$, $y$, and $z$ axes in crystals.}
	\label{tb_BCDBCQ}
	\begin{tabular}{l|l l l l} 
			\hline\hline
			& MPG & Generators & $D_{\alpha\beta}$ & $Q_{\alpha\beta\gamma}$
			\\ 
			\hline
			\multirow{13}{*}{Colorless}&
			$mm2$ & $C_{2z},M_y$ & $D_{xy},D_{yx}$
			& $Q_{xyz},Q_{xzy},Q_{yzx}$
			\\ &
			$mmm$ & $C_{2x},C_{2y},I$ &0 & $Q_{xyz},Q_{xzy},Q_{yzx}$
			\\
			&
			$4mm$ & $C_{4z},M_y$ & $D_{xy}=-D_{yx}$   &   $Q_{yzx}=-Q_{xzy}$  		
			\\  &
			$\bar{4}2m$ & $IC_{4z}, C_{2x}$ & $D_{xx}=-D_{yy} $   & 	$Q_{yzx}=-Q_{xzy}$
			\\  &
			$4/mmm$ & $C_{4z}, C_{2x}, I$ & 0 & 	 $Q_{yzx}=-Q_{xzy}$
			\\ 
			&
			$3m$ & $C_{3z},M_{x}$  &  $D_{xy}=-D_{yx}$    &   $\begin{gathered}[t]
				Q_{xxx}=-Q_{xyy}=-Q_{yyx},   Q_{yzx}=-Q_{xzy}
			\end{gathered}$  
			\\&
			$\bar{3}m$ & $C_{3z},C_{2x},I$  &0 &	$Q_{xxx}=-Q_{xyy}=-Q_{yyx},   Q_{yzx}=-Q_{xzy}$  
			\\ &
			$6mm$ & $C_{6z}, M_{y}$   & $D_{xy}=-D_{yx}$ &  $Q_{yzx}=-Q_{xzy}$  		
			\\  &
			$\bar{6}m2$ & $IC_{6z}, C_{2x}$ &0 &	$Q_{yzx}=-Q_{xzy}$
			\\  &
			$6/mmm$  & $C_{6z}, C_{2y}, I$ & 0 & 	$Q_{yzx}=-Q_{xzy}$
			\\  &
			$m\bar{3}$ & $	C_{2x}, C_{2y}, C_{3,111}, I$  & 0& $Q_{xyz}=Q_{yzx}=Q_{zxy}$ 
			\\ &
			$\bar{4}3m $ &  $\begin{gathered}[t]
				C_{2x}, C_{2y}, C_{3,111}, M_{110}
			\end{gathered}$  &  0 & 	0
			\\ &
			$m\bar{3}m$ & $ \begin{gathered}[t]
				C_{2x}, C_{2y}, C_{3,111},C_{2,110}, I
			\end{gathered}$  &0  & 0
			\\
			\hline \multirow{14}{*}{Black and white}&
			$\bar{4}^\prime$  & $TIC_{4z}$ & $\begin{gathered}[t]
				D_{xx} = -D_{yy}, D_{xy}= D_{yx}
			\end{gathered}$  & 	  $\begin{aligned}[t]
				&Q_{xxz} = -Q_{yyz}, Q_{xyz},\\ &Q_{xzx} = -Q_{yzy}, Q_{xzy} = Q_{yzx}
			\end{aligned}$  
			\\&
			$4^\prime/m$  & $TC_{4z},I$ & 0 &   $\begin{aligned}[t]
				&Q_{xxz} = -Q_{yyz}, Q_{xyz}, \\& Q_{xzx} = -Q_{yzy}, Q_{xzy} = Q_{yzx} \end{aligned}$  
			\\
			&
			$4^\prime m^\prime m $ & $TC_{4z}, M_{110}$ & $D_{xy}= -D_{yx}$
			& $Q_{xxz} = Q_{yyz}, Q_{xzx} = Q_{yzy}$
			\\  &
			$\bar{4}^\prime 2m^\prime$ & $TIC_{4z}, C_{2x}$ &  $D_{xx}=-D_{yy}$ &   $Q_{xyz}, Q_{xzy} = Q_{yzx}$
			\\  &
			$\bar{4}^\prime 2^\prime m$ & $TIC_{4z}, M_{110}$ & $D_{xy}= -D_{yx}$ &  $Q_{xxz} = Q_{yyz}, Q_{xzx} = Q_{yzy}$
			\\  &
			$4^\prime /mm^\prime m$ & $TC_{4z}, C_{2,110}, I$ & 0&   $Q_{xxz} = Q_{yyz}, Q_{xzx} = Q_{yzy}$
			\\
			&
			\textbf{$\bar{6}^\prime$} & $TIC_{6z}, C_{3z}$ &0  &  $\begin{aligned}[t] &Q_{xxx} = -Q_{xyy}=-Q_{yyx},\\ &Q_{xxy} =Q_{xyx}=- Q_{yyy} \end{aligned}$ 
			\\&
			\textbf{$6^\prime/m^\prime$}  & $TC_{6z},C_{3z},I$  &0   &  $\begin{aligned}[t] &Q_{xxx} = -Q_{xyy}=-Q_{yyx},\\ &Q_{xxy} =Q_{xyx}=- Q_{yyy} \end{aligned}$ 
			\\  &
			$6^\prime mm^\prime$ & $TC_{6z}, M_{y}$ & $D_{xy}= -D_{yx}$ &  $\begin{gathered}Q_{yyy} = -Q_{xxy}=-Q_{yxx}\end{gathered}$
			\\ & 
			$\bar{6}^\prime m^\prime2$ & $TIC_{6z}, C_{2x}$&0 &  $\begin{gathered}Q_{xxx} = -Q_{xyy}=-Q_{yyx}\end{gathered}$
			\\ & 
			$\bar{6}^\prime m2^\prime$ & $TIC_{6z}, M_y$ &0 &  $\begin{gathered}Q_{yyy} = -Q_{xxy}=-Q_{yxx}\end{gathered}$
			\\&  
			$6^\prime/m^\prime mm^\prime $ & $TC_{6z}, C_{2y}, I$&0 &  $\begin{gathered}Q_{yyy} = -Q_{xxy}=-Q_{yxx}\end{gathered}$
			\\  & 
			$\bar{4}^\prime3m^\prime$ & $\begin{gathered}[t]
				C_{2x}, C_{2y}, C_{3,111}, TM_{110}
			\end{gathered}$ & 0 &  $\begin{gathered}Q_{xyz} = Q_{yzx}=Q_{zxy}\end{gathered}$
			\\ & 
			$m\bar{3}m^{\prime}$ & $\begin{gathered}[t]
				C_{2x}, C_{2y}, C_{3,111},TC_{2,110}, I \end{gathered}$ 
			&0 & $\begin{gathered}Q_{xyz} = Q_{yzx}=Q_{zxy}\end{gathered}$
			\\ 
			\hline\hline
		\end{tabular}
\end{table*}
\section{Effect of nonsymmorphic operations}  
The MPG analysis in the main text identifies possible HNLs from the point-group symmetry constraints on the low-energy Hamiltonian. For symmorphic magnetic space groups, such as that of Mn$_3$GaN discussed in
the main text, this MPG analysis is sufficient to determine the HNL directions. For nonsymmorphic magnetic
space groups, a symmetry operation generally contains a fractional translation,
\(\tilde g=\{g|\boldsymbol{\tau}_g\}\), which introduces a
momentum-dependent phase factor in the representation matrix,
\begin{equation}
	D_{\mathbf k}(\tilde g)
	=
	e^{-i\mathbf k\cdot\boldsymbol{\tau}_g}D(g).
\end{equation}
Therefore, the full magnetic space group can modify the degeneracies predicted by the MPG analysis. Depending on the nonsymmorphic phase factor, an MPG-predicted HNL may be preserved, lifted, or supplemented by additional line
or plane degeneracies.  Compatibility relations of the full MSG can be used to verify the MPG-predicted HNLs. Although a general treatment of nonsymmorphic effects is case-dependent, we illustrate their possible roles below with
representative examples from the materials discussed in the main text.

One possible role of nonsymmorphic symmetries is to lift the HNL degeneracy predicted by the MPG analysis. This can be seen from the \(k_x\)-direction HNLs in MnTe. In its nonsymmorphic MSG \(Cmcm\), the mirror \(M_y\) and  \(M_z\) of the MPG
\(mmm\) are realized as the glide operations \(\tilde M_y=\{M_y|00\tfrac{1}{2}\}\) and \(\tilde M_z=\{M_z|00\tfrac{1}{2}\}\). 
Therefore, their representation matrices acquire a momentum-dependent phase $D_{\mathbf k}(\tilde M_y)
= e^{-ik_zc/2}D(M_y)$ and $D_{\mathbf k}(\tilde M_z)
= e^{-ik_zc/2}D(M_z)$. 
The algebra between \(\tilde M_y\) and \(\tilde M_z\) then becomes
\begin{equation}
	D_{\mathbf k}(\tilde M_y)D_{\mathbf k}(\tilde M_z)
	=
	-e^{-ik_zc}
	D_{\mathbf k}(\tilde M_z)D_{\mathbf k}(\tilde M_y).
\end{equation}
Here the minus sign represents the anticommutation of the two perpendicular pure mirrors \(M_y\) and \(M_z\), while the factor \(e^{-ik_zc}\) originates from the glide translation.
For the \(k_x\)-direction lines in the \(k_z=0\) plane, such as
\(\Gamma\)-X and Y-X\(_1\), the nonsymmorphic phase is trivial,
\(e^{-ik_zc}=1\). The two operators remain anticommuting, so the two-dimensional line corep is preserved and the MPG-predicted HNLs
survive. In contrast, for the \(k_x\)-direction lines in the \(k_z=\pi/c\)
plane, such as Z-A and T-A\(_1\), the phase factor becomes
\(e^{-ik_zc}=-1\), which cancels the spinful minus sign. The two operators then commute and can be simultaneously diagonalized. Consequently, the
two-dimensional line corep predicted by the MPG analysis splits
into two one-dimensional coreps, and the corresponding HNLs are
lifted. 
However, the HNLs along the other two directions are protected by the algebra involving the pure mirror $M_x=\{M_x|000\}$ and one glide mirror. 
In this case, the exchange of the two operations does not generate an additional Bloch phase, so the spinful anticommutation is preserved and the corresponding two-dimensional line corepresentations remain intact. 
For example, $D_{\mathbf k}(M_x)D_{\mathbf k}(\tilde M_z)
= -D_{\mathbf k}(\tilde M_z)D_{\mathbf k}(M_x)$.
 
  Another possible role of nonsymmorphic operations is to generate additional degeneracies at the Brillouin-zone boundary. This can be illustrated by CrSb, whose magnetic
  structure is described by the nonsymmorphic MSG \(P6_3'/m'm'c\). The
  antiunitary screw symmetry associated with \(\{ T C_{6z}|00\tfrac{1}{2}\}\) contains a fractional
  translation along the \(z\) direction. A convenient operation generated by this primed screw symmetry is \(\tilde{\mathcal A}
  = \{ T C_{2z}|00\tfrac{1}{2}\}\).  It acts on momentum as $\mathbf k  \rightarrow (k_x,k_y,-k_z)$. 
  Therefore, every momentum on the \(k_z=\pi/c\) plane is invariant under
  \(\tilde{\mathcal A}\) up to a reciprocal lattice vector. Acting twice with this antiunitary operation is equivalent to a full lattice translation along the \(z\) direction, and hence gives the Bloch phase
  \(e^{-ik_zc}\).
  On the Brillouin-zone-boundary plane \(k_z=\pi/c\),
  \begin{equation}
  	D_{k_z=\pi/c}(\tilde{\mathcal A})^2
  	=
  	e^{-i\pi}
  	=
  	-1 ,
  \end{equation}
  which enforces the Kramers-like twofold degeneracy in the \(k_z=\pi/c\) plane of CrSb.



\begin{thebibliography}{99}

\bibitem{Hasan_review_2010}
M.~Z. Hasan and C.~L. Kane.
\newblock {Colloquium: Topological insulators}.
\newblock {\em Rev. Mod. Phys.}, 82:3045--3067, Nov 2010.

\bibitem{Qi_review_2011}
Xiao-Liang Qi and Shou-Cheng Zhang.
\newblock {Topological insulators and superconductors}.
\newblock {\em Rev. Mod. Phys.}, 83:1057--1110, Oct 2011.

\bibitem{Armitage_review_2018}
N.~P. Armitage, E.~J. Mele, and Ashvin Vishwanath.
\newblock {Weyl and Dirac semimetals in three-dimensional solids}.
\newblock {\em Rev. Mod. Phys.}, 90:015001, Jan 2018.

\bibitem{PhysRevLett.108.140405}
S.~M. Young, S.~Zaheer, J.~C.~Y. Teo, C.~L. Kane, E.~J. Mele, and A.~M. Rappe.
\newblock {Dirac Semimetal in Three Dimensions}.
\newblock {\em Phys. Rev. Lett.}, 108:140405, Apr 2012.

\bibitem{PhysRevB.85.195320}
Zhijun Wang, Yan Sun, Xing-Qiu Chen, Cesare Franchini, Gang Xu, Hongming Weng, Xi~Dai, and Zhong Fang.
\newblock {Dirac semimetal and topological phase transitions in A$_{3}$Bi ($A=\text{Na}$, K, Rb)}.
\newblock {\em Phys. Rev. B}, 85:195320, May 2012.

\bibitem{PhysRevB.88.125427}
Zhijun Wang, Hongming Weng, Quansheng Wu, Xi~Dai, and Zhong Fang.
\newblock {Three-dimensional Dirac semimetal and quantum transport in Cd$_{3}$As$_{2}$}.
\newblock {\em Phys. Rev. B}, 88:125427, Sep 2013.

\bibitem{PhysRevLett.113.027603}
Sergey Borisenko, Quinn Gibson, Danil Evtushinsky, Volodymyr Zabolotnyy, Bernd B\"uchner, and Robert~J. Cava.
\newblock {Experimental Realization of a Three-Dimensional Dirac Semimetal}.
\newblock {\em Phys. Rev. Lett.}, 113:027603, Jul 2014.

\bibitem{10.1126/science.1245085}
Z.~K. Liu, B.~Zhou, Y.~Zhang, Z.~J. Wang, H.~M. Weng, D.~Prabhakaran, S.-K. Mo, Z.~X. Shen, Z.~Fang, X.~Dai, Z.~Hussain, and Y.~L. Chen.
\newblock {Discovery of a Three-Dimensional Topological Dirac Semimetal, Na$_3$Bi}.
\newblock {\em Science}, 343(6173):864--867, 2014.

\bibitem{Liu2014}
Z.~K. Liu, J.~Jiang, B.~Zhou, Z.~J. Wang, Y.~Zhang, H.~M. Weng, D.~Prabhakaran, S.-K. Mo, H.~Peng, P.~Dudin, T.~Kim, M.~Hoesch, Z.~Fang, X.~Dai, Z.~X. Shen, D.~L. Feng, Z.~Hussain, and Y.~L. Chen.
\newblock {A stable three-dimensional topological Dirac semimetal Cd$_3$As$_2$}.
\newblock {\em Nature Materials}, 13(7):677--681, Jul 2014.

\bibitem{Yang2014}
Bohm-Jung Yang and Naoto Nagaosa.
\newblock {Classification of stable three-dimensional Dirac semimetals with nontrivial topology}.
\newblock {\em Nature Communications}, 5(1):4898, Sep 2014.

\bibitem{10.1126/science.aac6089}
Jun Xiong, Satya~K. Kushwaha, Tian Liang, Jason~W. Krizan, Max Hirschberger, Wudi Wang, R.~J. Cava, and N.~P. Ong.
\newblock {Evidence for the chiral anomaly in the Dirac semimetal Na$_3$Bi }.
\newblock {\em Science}, 350(6259):413--416, 2015.

\bibitem{PhysRevB.83.205101}
Xiangang Wan, Ari~M. Turner, Ashvin Vishwanath, and Sergey~Y. Savrasov.
\newblock {Topological semimetal and Fermi-arc surface states in the electronic structure of pyrochlore iridates}.
\newblock {\em Phys. Rev. B}, 83:205101, May 2011.

\bibitem{PhysRevLett.107.127205}
A.~A. Burkov and Leon Balents.
\newblock {Weyl Semimetal in a Topological Insulator Multilayer}.
\newblock {\em Phys. Rev. Lett.}, 107:127205, Sep 2011.

\bibitem{PhysRevLett.107.186806}
Gang Xu, Hongming Weng, Zhijun Wang, Xi~Dai, and Zhong Fang.
\newblock {Chern Semimetal and the Quantized Anomalous Hall Effect in HgCr$_2$Se$_4$}.
\newblock {\em Phys. Rev. Lett.}, 107:186806, Oct 2011.

\bibitem{PhysRevB.84.075129}
Kai-Yu Yang, Yuan-Ming Lu, and Ying Ran.
\newblock {Quantum Hall effects in a Weyl semimetal: Possible application in pyrochlore iridates}.
\newblock {\em Phys. Rev. B}, 84:075129, Aug 2011.

\bibitem{PhysRevX.5.011029}
Hongming Weng, Chen Fang, Zhong Fang, B.~Andrei Bernevig, and Xi~Dai.
\newblock {Weyl Semimetal Phase in Noncentrosymmetric Transition-Metal Monophosphides}.
\newblock {\em Phys. Rev. X}, 5:011029, Mar 2015.

\bibitem{PhysRevX.5.031013}
B.~Q. Lv, H.~M. Weng, B.~B. Fu, X.~P. Wang, H.~Miao, J.~Ma, P.~Richard, X.~C. Huang, L.~X. Zhao, G.~F. Chen, Z.~Fang, X.~Dai, T.~Qian, and H.~Ding.
\newblock {Experimental Discovery of Weyl Semimetal TaAs}.
\newblock {\em Phys. Rev. X}, 5:031013, Jul 2015.

\bibitem{Xu2015}
Su-Yang Xu, Ilya Belopolski, Nasser Alidoust, Madhab Neupane, Guang Bian, Chenglong Zhang, Raman Sankar, Guoqing Chang, Zhujun Yuan, Chi-Cheng Lee, Shin-Ming Huang, Hao Zheng, Jie Ma, Daniel~S. Sanchez, BaoKai Wang, Arun Bansil, Fangcheng Chou, Pavel~P. Shibayev, Hsin Lin, Shuang Jia, and M.~Zahid Hasan.
\newblock {Discovery of a Weyl fermion semimetal and topological Fermi arcs}.
\newblock {\em Science}, 349(6248):613--617, 2015.

\bibitem{Wang2018}
Qi~Wang, Yuanfeng Xu, Rui Lou, Zhonghao Liu, Man Li, Yaobo Huang, Dawei Shen, Hongming Weng, Shancai Wang, and Hechang Lei.
\newblock {Large intrinsic anomalous Hall effect in half-metallic ferromagnet Co$_3$Sn$_2$S$_2$ with magnetic Weyl fermions}.
\newblock {\em Nature Communications}, 9(1):3681, Sep 2018.

\bibitem{Liu2019}
D.~F. Liu, A.~J. Liang, E.~K. Liu, Q.~N. Xu, Y.~W. Li, C.~Chen, D.~Pei, W.~J. Shi, S.~K. Mo, P.~Dudin, T.~Kim, C.~Cacho, G.~Li, Y.~Sun, L.~X. Yang, Z.~K. Liu, S.~S.~P. Parkin, C.~Felser, and Y.~L. Chen.
\newblock {Magnetic Weyl semimetal phase in a Kagomé crystal}.
\newblock {\em Science}, 365(6459):1282--1285, 2019.

\bibitem{Noam2019}
Noam Morali, Rajib Batabyal, Pranab~Kumar Nag, Enke Liu, Qiunan Xu, Yan Sun, Binghai Yan, Claudia Felser, Nurit Avraham, and Haim Beidenkopf.
\newblock {Fermi-arc diversity on surface terminations of the magnetic Weyl semimetal Co$_3$Sn$_2$S$_2$}.
\newblock {\em Science}, 365(6459):1286--1291, 2019.

\bibitem{PhysRevB.84.235126}
A.~A. Burkov, M.~D. Hook, and Leon Balents.
\newblock {Topological nodal semimetals}.
\newblock {\em Phys. Rev. B}, 84:235126, Dec 2011.

\bibitem{PhysRevB.90.205136}
Ching-Kai Chiu and Andreas~P. Schnyder.
\newblock {Classification of reflection-symmetry-protected topological semimetals and nodal superconductors}.
\newblock {\em Phys. Rev. B}, 90:205136, Nov 2014.

\bibitem{PhysRevB.92.045108}
Hongming Weng, Yunye Liang, Qiunan Xu, Rui Yu, Zhong Fang, Xi~Dai, and Yoshiyuki Kawazoe.
\newblock {Topological node-line semimetal in three-dimensional graphene networks}.
\newblock {\em Phys. Rev. B}, 92:045108, Jul 2015.

\bibitem{PhysRevLett.115.036806}
Youngkuk Kim, Benjamin~J. Wieder, C.~L. Kane, and Andrew~M. Rappe.
\newblock {Dirac Line Nodes in Inversion-Symmetric Crystals}.
\newblock {\em Phys. Rev. Lett.}, 115:036806, Jul 2015.

\bibitem{PhysRevB.92.081201}
Chen Fang, Yige Chen, Hae-Young Kee, and Liang Fu.
\newblock {Topological nodal line semimetals with and without spin-orbital coupling}.
\newblock {\em Phys. Rev. B}, 92:081201, Aug 2015.

\bibitem{Bian2016}
Guang Bian, Tay-Rong Chang, Raman Sankar, Su-Yang Xu, Hao Zheng, Titus Neupert, Ching-Kai Chiu, Shin-Ming Huang, Guoqing Chang, Ilya Belopolski, Daniel~S. Sanchez, Madhab Neupane, Nasser Alidoust, Chang Liu, BaoKai Wang, Chi-Cheng Lee, Horng-Tay Jeng, Chenglong Zhang, Zhujun Yuan, Shuang Jia, Arun Bansil, Fangcheng Chou, Hsin Lin, and M.~Zahid Hasan.
\newblock {Topological nodal-line fermions in spin-orbit metal PbTaSe$_2$}.
\newblock {\em Nature Communications}, 7(1):10556, Feb 2016.

\bibitem{Schoop2016}
Leslie~M. Schoop, Mazhar~N. Ali, Carola Stra{\ss}er, Andreas Topp, Andrei Varykhalov, Dmitry Marchenko, Viola Duppel, Stuart S.~P. Parkin, Bettina~V. Lotsch, and Christian~R. Ast.
\newblock {Dirac cone protected by non-symmorphic symmetry and three-dimensional Dirac line node in ZrSiS}.
\newblock {\em Nature Communications}, 7(1):11696, May 2016.

\bibitem{Wang2017}
Jing Wang.
\newblock {Antiferromagnetic topological nodal line semimetals}.
\newblock {\em Phys. Rev. B}, 96:081107(R), Aug 2017.

\bibitem{fu2019}
B.-B. Fu, C.-J. Yi, T.-T. Zhang, M.~Caputo, J.-Z. Ma, X.~Gao, B.~Q. Lv, L.-Y. Kong, Y.-B. Huang, P.~Richard, M.~Shi, V.~N. Strocov, C.~Fang, H.-M. Weng, Y.-G. Shi, T.~Qian, and H.~Ding.
\newblock {Dirac nodal surfaces and nodal lines in ZrSiS}.
\newblock {\em Science Advances}, 5(5):eaau6459, 2019.

\bibitem{Ilya2019}
Ilya Belopolski, Kaustuv Manna, Daniel~S. Sanchez, Guoqing Chang, Benedikt Ernst, Jiaxin Yin, Songtian~S. Zhang, Tyler Cochran, Nana Shumiya, Hao Zheng, Bahadur Singh, Guang Bian, Daniel Multer, Maksim Litskevich, Xiaoting Zhou, Shin-Ming Huang, Baokai Wang, Tay-Rong Chang, Su-Yang Xu, Arun Bansil, Claudia Felser, Hsin Lin, and M.~Zahid Hasan.
\newblock {Discovery of topological Weyl fermion lines and drumhead surface states in a room temperature magnet}.
\newblock {\em Science}, 365(6459):1278--1281, 2019.

\bibitem{Yang2021}
Zeying Zhang, Zhi-Ming Yu, and Shengyuan~A. Yang.
\newblock {Magnetic higher-order nodal lines}.
\newblock {\em Phys. Rev. B}, 103:115112, Mar 2021.

\bibitem{Knoll2022}
Andy Knoll and Carsten Timm.
\newblock {Classification of Weyl points and nodal lines based on magnetic point groups for spin-$\frac{1}{2}$ quasiparticles}.
\newblock {\em Phys. Rev. B}, 105:115109, Mar 2022.

\bibitem{Zhuang2026}
Zheng-Yang Zhuang, Di~Zhu, Zhigang Wu, and Zhongbo Yan.
\newblock {Cartesian nodal lines and magnetic Kramers Weyl nodes in spin-split antiferromagnets}.
\newblock {\em Newton}, 2(7):100403, Jul 2026.

\bibitem{Nielsen1983}
H.B. Nielsen and Masao Ninomiya.
\newblock {The Adler-Bell-Jackiw anomaly and Weyl fermions in a crystal}.
\newblock {\em Physics Letters B}, 130(6):389--396, 1983.

\bibitem{Son2013}
D.~T. Son and B.~Z. Spivak.
\newblock {Chiral anomaly and classical negative magnetoresistance of Weyl metals}.
\newblock {\em Phys. Rev. B}, 88:104412, Sep 2013.

\bibitem{Huang2015}
Xiaochun Huang, Lingxiao Zhao, Yujia Long, Peipei Wang, Dong Chen, Zhanhai Yang, Hui Liang, Mianqi Xue, Hongming Weng, Zhong Fang, Xi~Dai, and Genfu Chen.
\newblock {Observation of the Chiral-Anomaly-Induced Negative Magnetoresistance in 3D Weyl Semimetal TaAs}.
\newblock {\em Phys. Rev. X}, 5:031023, Aug 2015.

\bibitem{Zhang2016}
Cheng-Long Zhang, Su-Yang Xu, Ilya Belopolski, Zhujun Yuan, Ziquan Lin, Bingbing Tong, Guang Bian, Nasser Alidoust, Chi-Cheng Lee, Shin-Ming Huang, Tay-Rong Chang, Guoqing Chang, Chuang-Han Hsu, Horng-Tay Jeng, Madhab Neupane, Daniel~S. Sanchez, Hao Zheng, Junfeng Wang, Hsin Lin, Chi Zhang, Hai-Zhou Lu, Shun-Qing Shen, Titus Neupert, M.~Zahid~Hasan, and Shuang Jia.
\newblock {Signatures of the Adler--Bell--Jackiw chiral anomaly in a Weyl fermion semimetal}.
\newblock {\em Nature Communications}, 7(1):10735, Feb 2016.

\bibitem{Chan2016}
Ching-Kit Chan, Patrick~A. Lee, Kenneth~S. Burch, Jung~Hoon Han, and Ying Ran.
\newblock {When Chiral Photons Meet Chiral Fermions: Photoinduced Anomalous Hall Effects in Weyl Semimetals}.
\newblock {\em Phys. Rev. Lett.}, 116:026805, Jan 2016.

\bibitem{deJuan2017}
Fernando de~Juan, Adolfo~G. Grushin, Takahiro Morimoto, and Joel~E. Moore.
\newblock {Quantized circular photogalvanic effect in Weyl semimetals}.
\newblock {\em Nature Communications}, 8(1):15995, Jul 2017.

\bibitem{Wu2017}
Liang Wu, S.~Patankar, T.~Morimoto, N.~L. Nair, E.~Thewalt, A.~Little, J.~G. Analytis, J.~E. Moore, and J.~Orenstein.
\newblock {Giant anisotropic nonlinear optical response in transition metal monopnictide Weyl semimetals}.
\newblock {\em Nature Physics}, 13(4):350--355, Apr 2017.

\bibitem{Ashby2013}
Phillip E.~C. Ashby and J.~P. Carbotte.
\newblock {Magneto-optical conductivity of Weyl semimetals}.
\newblock {\em Phys. Rev. B}, 87:245131, Jun 2013.

\bibitem{Kargarian2015}
Mehdi Kargarian, Mohit Randeria, and Nandini Trivedi.
\newblock {Theory of Kerr and Faraday rotations and linear dichroism in Topological Weyl Semimetals}.
\newblock {\em Scientific Reports}, 5(1):12683, Aug 2015.

\bibitem{Okamura2020}
Y.~Okamura, S.~Minami, Y.~Kato, Y.~Fujishiro, Y.~Kaneko, J.~Ikeda, J.~Muramoto, R.~Kaneko, K.~Ueda, V.~Kocsis, N.~Kanazawa, Y.~Taguchi, T.~Koretsune, K.~Fujiwara, A.~Tsukazaki, R.~Arita, Y.~Tokura, and Y.~Takahashi.
\newblock {Giant magneto-optical responses in magnetic Weyl semimetal Co$_3$Sn$_2$S$_2$}.
\newblock {\em Nature Communications}, 11(1):4619, Sep 2020.

\bibitem{Po2017}
Hoi~Chun Po, Ashvin Vishwanath, and Haruki Watanabe.
\newblock {Symmetry-based indicators of band topology in the 230 space groups}.
\newblock {\em Nature Communications}, 8(1):50, Jun 2017.

\bibitem{PKruthoff2017}
Jorrit Kruthoff, Jan de~Boer, Jasper van Wezel, Charles~L. Kane, and Robert-Jan Slager.
\newblock {Topological Classification of Crystalline Insulators through Band Structure Combinatorics}.
\newblock {\em Phys. Rev. X}, 7:041069, Dec 2017.

\bibitem{Watanabe2018}
Haruki Watanabe, Hoi~Chun Po, and Ashvin Vishwanath.
\newblock {Structure and topology of band structures in the 1651 magnetic space groups}.
\newblock {\em Science Advances}, 4(8):eaat8685, 2018.

\bibitem{Elcoro2021}
Luis Elcoro, Benjamin~J. Wieder, Zhida Song, Yuanfeng Xu, Barry Bradlyn, and B.~Andrei Bernevig.
\newblock {Magnetic topological quantum chemistry}.
\newblock {\em Nature Communications}, 12(1):5965, Oct 2021.

\bibitem{Chang2018}
Guoqing Chang, Benjamin~J. Wieder, Frank Schindler, Daniel~S. Sanchez, Ilya Belopolski, Shin-Ming Huang, Bahadur Singh, Di~Wu, Tay-Rong Chang, Titus Neupert, Su-Yang Xu, Hsin Lin, and M.~Zahid Hasan.
\newblock {Topological quantum properties of chiral crystals}.
\newblock {\em Nature Materials}, 17(11):978--985, Nov 2018.

\bibitem{Xie2021}
Ying-Ming Xie, Xue-Jian Gao, Xiao~Yan Xu, Cheng-Ping Zhang, Jin-Xin Hu, Jason~Z. Gao, and K.~T. Law.
\newblock {Kramers nodal line metals}.
\newblock {\em Nature Communications}, 12(1):3064, May 2021.

\bibitem{gao2023}
Xue-Jian Gao, Zi-Ting Sun, Ruo-Peng Yu, Xing-Yao Guo, and K.~T. Law.
\newblock {Heesch Weyl Fermions in inadmissible chiral antiferromagnets}, 2023.
\newblock arXiv:2305.15876.

\bibitem{Zhang2023}
Yichen Zhang, Yuxiang Gao, Xue-Jian Gao, Shiming Lei, Zhuoliang Ni, Ji~Seop Oh, Jianwei Huang, Ziqin Yue, Marta Zonno, Sergey Gorovikov, Makoto Hashimoto, Donghui Lu, Jonathan~D. Denlinger, Robert~J. Birgeneau, Junichiro Kono, Liang Wu, Kam~Tuen Law, Emilia Morosan, and Ming Yi.
\newblock {Kramers nodal lines and Weyl fermions in SmAlSi}.
\newblock {\em Communications Physics}, 6(1):134, Jun 2023.

\bibitem{Sarkar2023}
Shuvam Sarkar, Joydipto Bhattacharya, Pampa Sadhukhan, Davide Curcio, Rajeev Dutt, Vipin~Kumar Singh, Marco Bianchi, Arnab Pariari, Shubhankar Roy, Prabhat Mandal, Tanmoy Das, Philip Hofmann, Aparna Chakrabarti, and Sudipta Roy~Barman.
\newblock {Charge density wave induced nodal lines in LaTe$_3$}.
\newblock {\em Nature Communications}, 14(1):3628, Jun 2023.

\bibitem{Kurumaji2025}
Takashi Kurumaji, Jorge~I. Facio, Natsuki Mitsuishi, Shusaku Imajo, Masaki Gen, Motoi Kimata, Linda Ye, David Graf, Masato Sakano, Miho Kitamura, Kohei Yamagami, Kyoko Ishizaka, Koichi Kindo, and Taka-hisa Arima.
\newblock {Electronic Structure of Kramers Nodal-Line Semimetal YAuGe and Anomalous Hall Effect Induced by Magnetic Rare-Earth Substitution}.
\newblock {\em Advanced Science}, 12(27):2501669, 2025.

\bibitem{Zhang2025}
Yichen Zhang, Yuxiang Gao, Aki Pulkkinen, Xingyao Guo, Jianwei Huang, Yucheng Guo, Ziqin Yue, Ji~Seop Oh, Alex Moon, Mohamed Oudah, Xue-Jian Gao, Alberto Marmodoro, Alexei Fedorov, Sung-Kwan Mo, Makoto Hashimoto, Donghui Lu, Anil Rajapitamahuni, Elio Vescovo, Junichiro Kono, Alannah~M. Hallas, Robert~J. Birgeneau, Luis Balicas, J{\'a}n Min{\'a}r, Pavan Hosur, Kam~Tuen Law, Emilia Morosan, and Ming Yi.
\newblock {Kramers nodal lines in intercalated TaS$_2$ superconductors}.
\newblock {\em Nature Communications}, 16(1):4984, May 2025.

\bibitem{Gabriele2025}
Gabriele Domaine, Moritz~M. Hirschmann, Kirill Parshukov, Mihir Date, Holger~L. Meyerheim, Matthew~D. Watson, Katayoon Mohseni, Sydney K.~Y. Dufresne, Shigemi Terakawa, Marcin Rosmus, Natalia Olszowska, Stuart S.~P. Parkin, Andreas~P. Schnyder, and Niels B.~M. Schr{\"o}ter.
\newblock {Tunable Octdong and Spindle-Torus Fermi Surfaces in Kramers Nodal Line Metals}.
\newblock {\em Nature Communications}, 16(1):11128, Dec 2025.

\bibitem{Sarkar2026}
Shuvam Sarkar, Joydipto Bhattacharya, Pramod Bhakuni, Divya Jangra, Pampa Sadhukhan, Rajib Batabyal, Christos~D. Malliakas, Marco Bianchi, Davide Curcio, Shubhankar Roy, Arnab Pariari, Sajal Barman, Mohammad Balal, Giovanni Di~Santo, Luca Petaccia, Duck~Young Chung, Yihao Wang, Vasant~G. Sathe, Prabhat Mandal, Mercouri~G. Kanatzidis, Philip Hofmann, Aparna Chakrabarti, and Sudipta~Roy Barman.
\newblock {Kramers nodal line in the charge density wave state of ${\mathrm{YTe}}_{3}$ and the influence of twin domains}.
\newblock {\em Phys. Rev. B}, 113:035129, Jan 2026.

\bibitem{Bilbao}
Luis Elcoro, Barry Bradlyn, Zhijun Wang, Maia~G. Vergniory, Jennifer Cano, Claudia Felser, B.~Andrei Bernevig, Danel Orobengoa, Gemma de~la Flor, and Mois~I. Aroyo.
\newblock {Double crystallographic groups and their representations on the Bilbao Crystallographic Server}.
\newblock {\em Journal of Applied Crystallography}, 50(5):1457--1477, Oct 2017.

\bibitem{MAGNDATA_I}
Samuel~V. Gallego, J.~Manuel Perez-Mato, Luis Elcoro, Emre~S. Tasci, Robert~M. Hanson, Koichi Momma, Mois~I. Aroyo, and Gotzon Madariaga.
\newblock {{\it MAGNDATA}: towards a database of magnetic structures. I. The commensurate case}.
\newblock {\em Journal of Applied Crystallography}, 49(5):1750--1776, Oct 2016.

\bibitem{MAGNDATA_II}
Samuel~V. Gallego, J.~Manuel Perez-Mato, Luis Elcoro, Emre~S. Tasci, Robert~M. Hanson, Mois~I. Aroyo, and Gotzon Madariaga.
\newblock {{\it MAGNDATA}: towards a database of magnetic structures. II. The incommensurate case}.
\newblock {\em Journal of Applied Crystallography}, 49(6):1941--1956, Dec 2016.

\bibitem{book_group}
C.~J. Bradley and A.~P. Cracknell.
\newblock {\em {The Mathematical Theory Of Symmetry In Solids: Representation theory for point groups and space groups}}.
\newblock Oxford University Press, 2009.

\bibitem{PhysRevLett.133.206401}
Jianyang Ding, Zhicheng Jiang, Xiuhua Chen, Zicheng Tao, Zhengtai Liu, Tongrui Li, Jishan Liu, Jianping Sun, Jinguang Cheng, Jiayu Liu, Yichen Yang, Runfeng Zhang, Liwei Deng, Wenchuan Jing, Yu~Huang, Yuming Shi, Mao Ye, Shan Qiao, Yilin Wang, Yanfeng Guo, Donglai Feng, and Dawei Shen.
\newblock {Large Band Splitting in $g$-Wave Altermagnet CrSb}.
\newblock {\em Phys. Rev. Lett.}, 133:206401, Nov 2024.

\bibitem{Reimers2024}
Sonka Reimers, Lukas Odenbreit, Libor {\v{S}}mejkal, Vladimir~N. Strocov, Procopios Constantinou, Anna~B. Hellenes, Rodrigo Jaeschke~Ubiergo, Warlley~H. Campos, Venkata~K. Bharadwaj, Atasi Chakraborty, Thibaud Denneulin, Wen Shi, Rafal~E. Dunin-Borkowski, Suvadip Das, Mathias Kl{\"a}ui, Jairo Sinova, and Martin Jourdan.
\newblock {Direct observation of altermagnetic band splitting in CrSb thin films}.
\newblock {\em Nature Communications}, 15(1):2116, Mar 2024.

\bibitem{Yang2025}
Guowei Yang, Zhanghuan Li, Sai Yang, Jiyuan Li, Hao Zheng, Weifan Zhu, Ze~Pan, Yifu Xu, Saizheng Cao, Wenxuan Zhao, Anupam Jana, Jiawen Zhang, Mao Ye, Yu~Song, Lun-Hui Hu, Lexian Yang, Jun Fujii, Ivana Vobornik, Ming Shi, Huiqiu Yuan, Yongjun Zhang, Yuanfeng Xu, and Yang Liu.
\newblock {Three-dimensional mapping of the altermagnetic spin splitting in CrSb}.
\newblock {\em Nature Communications}, 16(1):1442, Feb 2025.

\bibitem{PhysRevB.109.024404}
Rafael~M. Fernandes, Vanuildo~S. de~Carvalho, Turan Birol, and Rodrigo~G. Pereira.
\newblock {Topological transition from nodal to nodeless Zeeman splitting in altermagnets}.
\newblock {\em Phys. Rev. B}, 109:024404, Jan 2024.

\bibitem{Smejkal2022}
Libor {\v{S}}mejkal, Allan~H. MacDonald, Jairo Sinova, Satoru Nakatsuji, and Tomas Jungwirth.
\newblock {Anomalous Hall antiferromagnets}.
\newblock {\em Nature Reviews Materials}, 7(6):482--496, Jun 2022.

\bibitem{PhysRevLett.115.216806}
Inti Sodemann and Liang Fu.
\newblock {Quantum Nonlinear Hall Effect Induced by Berry Curvature Dipole in Time-Reversal Invariant Materials}.
\newblock {\em Phys. Rev. Lett.}, 115:216806, Nov 2015.

\bibitem{PhysRevB.107.115142}
Cheng-Ping Zhang, Xue-Jian Gao, Ying-Ming Xie, Hoi~Chun Po, and K.~T. Law.
\newblock {Higher-order nonlinear anomalous Hall effects induced by Berry curvature multipoles}.
\newblock {\em Phys. Rev. B}, 107:115142, Mar 2023.

\bibitem{PhysRevB.62.R6065}
Kenya Ohgushi, Shuichi Murakami, and Naoto Nagaosa.
\newblock {Spin anisotropy and quantum Hall effect in the kagom\'e lattice: Chiral spin state based on a ferromagnet}.
\newblock {\em Phys. Rev. B}, 62:R6065--R6068, Sep 2000.

\bibitem{Chen2014}
Hua Chen, Qian Niu, and A.~H. MacDonald.
\newblock {Anomalous Hall Effect Arising from Noncollinear Antiferromagnetism}.
\newblock {\em Phys. Rev. Lett.}, 112:017205, Jan 2014.

\bibitem{Sancho_1984}
M~P~Lopez Sancho, J~M~Lopez Sancho, and J~Rubio.
\newblock {Quick iterative scheme for the calculation of transfer matrices: application to Mo (100)}.
\newblock {\em Journal of Physics F: Metal Physics}, 14(5):1205, May 1984.

\bibitem{Sancho_1985}
M~P~Lopez Sancho, J~M~Lopez Sancho, J~M~L Sancho, and J~Rubio.
\newblock {Highly convergent schemes for the calculation of bulk and surface Green functions}.
\newblock {\em Journal of Physics F: Metal Physics}, 15(4):851, Apr 1985.

\bibitem{Niu2014}
Yang Gao, Shengyuan~A. Yang, and Qian Niu.
\newblock {Field Induced Positional Shift of Bloch Electrons and Its Dynamical Implications}.
\newblock {\em Phys. Rev. Lett.}, 112:166601, Apr 2014.

\bibitem{Di2021}
Chong Wang, Yang Gao, and Di~Xiao.
\newblock {Intrinsic Nonlinear Hall Effect in Antiferromagnetic Tetragonal CuMnAs}.
\newblock {\em Phys. Rev. Lett.}, 127:277201, Dec 2021.

\bibitem{Yang2021_qmd}
Huiying Liu, Jianzhou Zhao, Yue-Xin Huang, Weikang Wu, Xian-Lei Sheng, Cong Xiao, and Shengyuan~A. Yang.
\newblock {Intrinsic Second-Order Anomalous Hall Effect and Its Application in Compensated Antiferromagnets}.
\newblock {\em Phys. Rev. Lett.}, 127:277202, Dec 2021.

\bibitem{Das2023}
Kamal Das, Shibalik Lahiri, Rhonald~Burgos Atencia, Dimitrie Culcer, and Amit Agarwal.
\newblock {Intrinsic nonlinear conductivities induced by the quantum metric}.
\newblock {\em Phys. Rev. B}, 108:L201405, Nov 2023.

\bibitem{Yan2024}
Daniel Kaplan, Tobias Holder, and Binghai Yan.
\newblock {Unification of Nonlinear Anomalous Hall Effect and Nonreciprocal Magnetoresistance in Metals by the Quantum Geometry}.
\newblock {\em Phys. Rev. Lett.}, 132:026301, Jan 2024.

\bibitem{Suyang2023}
Anyuan Gao, Yu-Fei Liu, Jian-Xiang Qiu, Barun Ghosh, Thaís~V. Trevisan, Yugo Onishi, Chaowei Hu, Tiema Qian, Hung-Ju Tien, Shao-Wen Chen, Mengqi Huang, Damien Bérubé, Houchen Li, Christian Tzschaschel, Thao Dinh, Zhe Sun, Sheng-Chin Ho, Shang-Wei Lien, Bahadur Singh, Kenji Watanabe, Takashi Taniguchi, David~C. Bell, Hsin Lin, Tay-Rong Chang, Chunhui~Rita Du, Arun Bansil, Liang Fu, Ni~Ni, Peter~P. Orth, Qiong Ma, and Su-Yang Xu.
\newblock {Quantum metric nonlinear Hall effect in a topological antiferromagnetic heterostructure}.
\newblock {\em Science}, 381(6654):181--186, 2023.

\bibitem{Wang2023}
Naizhou Wang, Daniel Kaplan, Zhaowei Zhang, Tobias Holder, Ning Cao, Aifeng Wang, Xiaoyuan Zhou, Feifei Zhou, Zhengzhi Jiang, Chusheng Zhang, Shihao Ru, Hongbing Cai, Kenji Watanabe, Takashi Taniguchi, Binghai Yan, and Weibo Gao.
\newblock {Quantum-metric-induced nonlinear transport in a topological antiferromagnet}.
\newblock {\em Nature}, 621(7979):487--492, Sep 2023.

\bibitem{Zhao2025}
Tong-Yang Zhao, An-Qi Wang, Zhen-Tao Zhang, Zheng-Yang Cao, Xing-Yu Liu, and Zhi-Min Liao.
\newblock {Magnetic Field Induced Quantum Metric Dipole in Dirac Semimetal Cd$_{3}$As$_{2}$}.
\newblock {\em Phys. Rev. Lett.}, 135:026601, Jul 2025.

\bibitem{Samathrakis2020}
Ilias Samathrakis and Hongbin Zhang.
\newblock {Tailoring the anomalous Hall effect in the noncollinear antiperovskite ${\mathrm{Mn}}_{3}\mathrm{GaN}$}.
\newblock {\em Phys. Rev. B}, 101:214423, Jun 2020.

\bibitem{Zhou2020}
Xiaodong Zhou, Jan-Philipp Hanke, Wanxiang Feng, Stefan Bl\"ugel, Yuriy Mokrousov, and Yugui Yao.
\newblock {Giant anomalous Nernst effect in noncollinear antiferromagnetic Mn-based antiperovskite nitrides}.
\newblock {\em Phys. Rev. Mater.}, 4:024408, Feb 2020.

\bibitem{sun2025}
Yu-Hao Wan, Peng-Yi Liu, and Qing-Feng Sun.
\newblock {Quantum Anomalous Hall Effect in Ferromagnetic Metals}.
\newblock {\em Phys. Rev. Lett.}, 135:186302, Oct 2025.

\bibitem{Kübler_2014}
J.~Kübler and C.~Felser.
\newblock {Non-collinear antiferromagnets and the anomalous Hall effect}.
\newblock {\em Europhysics Letters}, 108(6):67001, Dec 2014.

\bibitem{Nakatsuji2015}
Satoru Nakatsuji, Naoki Kiyohara, and Tomoya Higo.
\newblock {Large anomalous Hall effect in a non-collinear antiferromagnet at room temperature}.
\newblock {\em Nature}, 527(7577):212--215, Nov 2015.

\bibitem{Ajaya2016}
Ajaya~K. Nayak, Julia~Erika Fischer, Yan Sun, Binghai Yan, Julie Karel, Alexander~C. Komarek, Chandra Shekhar, Nitesh Kumar, Walter Schnelle, Jürgen Kübler, Claudia Felser, and Stuart S.~P. Parkin.
\newblock {Large anomalous Hall effect driven by a nonvanishing Berry curvature in the noncolinear antiferromagnet Mn$_3$Ge}.
\newblock {\em Science Advances}, 2(4):e1501870, 2016.

\bibitem{Yan2017}
Yang Zhang, Yan Sun, Hao Yang, Jakub \ifmmode~\check{Z}\else \v{Z}\fi{}elezn\'y, Stuart P.~P. Parkin, Claudia Felser, and Binghai Yan.
\newblock {Strong anisotropic anomalous Hall effect and spin Hall effect in the chiral antiferromagnetic compounds Mn$_{3}$X ($X=\mathrm{Ge}$, Sn, Ga, Ir, Rh, and Pt)}.
\newblock {\em Phys. Rev. B}, 95:075128, Feb 2017.

\bibitem{Gurung2019}
Gautam Gurung, Ding-Fu Shao, Tula~R. Paudel, and Evgeny~Y. Tsymbal.
\newblock {Anomalous Hall conductivity of noncollinear magnetic antiperovskites}.
\newblock {\em Phys. Rev. Mater.}, 3:044409, Apr 2019.

\end{thebibliography}

\begin{thebibliography}{1}

\bibitem{SM_KRESSE199615}
G.~Kresse and J.~Furthmüller.
\newblock {Efficiency of ab-initio total energy calculations for metals and semiconductors using a plane-wave basis set}.
\newblock {\em Computational Materials Science}, 6(1):15--50, 1996.

\bibitem{SM_PhysRevB.50.17953}
P.~E. Bl\"ochl.
\newblock {Projector augmented-wave method}.
\newblock {\em Phys. Rev. B}, 50:17953--17979, Dec 1994.

\bibitem{SM_PhysRevB.28.1809}
David~C. Langreth and M.~J. Mehl.
\newblock {Beyond the local-density approximation in calculations of ground-state electronic properties}.
\newblock {\em Phys. Rev. B}, 28:1809--1834, Aug 1983.

\bibitem{SM_PhysRevLett.77.3865}
John~P. Perdew, Kieron Burke, and Matthias Ernzerhof.
\newblock {Generalized Gradient Approximation Made Simple}.
\newblock {\em Phys. Rev. Lett.}, 77:3865--3868, Oct 1996.

\bibitem{SM_PhysRev.136.B864}
P.~Hohenberg and W.~Kohn.
\newblock {Inhomogeneous Electron Gas}.
\newblock {\em Phys. Rev.}, 136:B864--B871, Nov 1964.

\bibitem{SM_MAGNDATA_I}
Samuel~V. Gallego, J.~Manuel Perez-Mato, Luis Elcoro, Emre~S. Tasci, Robert~M. Hanson, Koichi Momma, Mois~I. Aroyo, and Gotzon Madariaga.
\newblock {{\it MAGNDATA}: towards a database of magnetic structures. I. The commensurate case}.
\newblock {\em Journal of Applied Crystallography}, 49(5):1750--1776, Oct 2016.

\bibitem{SM_MAGNDATA_II}
Samuel~V. Gallego, J.~Manuel Perez-Mato, Luis Elcoro, Emre~S. Tasci, Robert~M. Hanson, Mois~I. Aroyo, and Gotzon Madariaga.
\newblock {{\it MAGNDATA}: towards a database of magnetic structures. II. The incommensurate case}.
\newblock {\em Journal of Applied Crystallography}, 49(6):1941--1956, Dec 2016.

\end{thebibliography}
\end{document}